\documentclass[
journal=ancac3,
manuscript=article,
layout=traditional,
email=true,
]{achemso}
\setkeys{acs}{maxauthors=4}
\setkeys{acs}{doi=true}

\PassOptionsToPackage{italian,english}{babel}               
    \usepackage{babel}
\usepackage{silence}
\usepackage{graphicx}                                       
\usepackage{dcolumn}                                        
\usepackage[mathlines]{lineno}                              
\usepackage{xcolor}                                         
\usepackage{float}                                          
\usepackage{tabularx}                                       
\usepackage[section]{placeins}								
\usepackage[version=4]{mhchem}								
\usepackage{amsmath}
\usepackage{amsfonts}
\usepackage{amssymb}
\usepackage{bm}                                             
\usepackage{hyperref}                                       
\usepackage{epsfig}
\usepackage{pdfpages}
\usepackage{pdflscape}
\usepackage{doi}

\usepackage{xr}
\hypersetup{%
  colorlinks=true, linktocpage=true, pdfstartpage=1, pdfstartview=FitV,%
  breaklinks=true, pageanchor=true,%
  pdfpagemode=UseNone, %
  plainpages=false, bookmarksnumbered, bookmarksopen=true, bookmarksopenlevel=1,%
  hypertexnames=true, pdfhighlight=/O,
  urlcolor=cyan, linkcolor=orange, citecolor=red, 
  pdftitle={Ultra-thin transistors and circuits for conformable electronics},%
  pdfauthor={Federico Parenti}%
}
\makeatletter
\let\cat@comma@active\@empty
\makeatother

\title{Ultra-thin transistors and circuits for conformable electronics} 
\author{Federico Parenti}
\affiliation{Dipartimento di Ingegneria dell' Informazione - Università di Pisa, I-56122, Pisa, Italy}
\author{Riccardo Sargeni}
\affiliation{Dipartimento di Ingegneria dell' Informazione - Università di Pisa, I-56122, Pisa, Italy}
\alsoaffiliation{Quantavis s.r.l., Largo Padre Renzo Spadoni, I-56126 Pisa, Italy}
\author{Elisabetta Dimaggio}
\affiliation{Dipartimento di Ingegneria dell' Informazione - Università di Pisa, I-56122, Pisa, Italy}
\author{Francesco Pieri}
\affiliation{Dipartimento di Ingegneria dell' Informazione - Università di Pisa, I-56122, Pisa, Italy}
\author{Filippo Fabbri}
\affiliation{NEST Laboratory - Istituto Nanoscienze-CNR and Scuola Normale Superiore, I-56127, Pisa, Italy}
\author{Tommaso Losi}
\affiliation{Center for Nano Science and Technology - Istituto Italiano di Tecnologia (IIT), I-20134, Milano, Italy}
\author{Fabrizio Antonio Viola}
\affiliation{Center for Nano Science and Technology - Istituto Italiano di Tecnologia (IIT), I-20134, Milano, Italy}
\author{Arindam Bala}
\affiliation{Institute of Electrical and Microengineering - École Polytechnique Fédérale de Lausanne (EPFL), CH-1015, Lausanne, Switzerland}
\author{Zhenyu Wang}
\affiliation{Institute of Electrical and Microengineering - École Polytechnique Fédérale de Lausanne (EPFL), CH-1015, Lausanne, Switzerland}
\author{Andras Kis}
\affiliation{Institute of Electrical and Microengineering - École Polytechnique Fédérale de Lausanne (EPFL), CH-1015, Lausanne, Switzerland}
\author{Mario Caironi}
\affiliation{Center for Nano Science and Technology - Istituto Italiano di Tecnologia (IIT), I-20134, Milano, Italy}
\author{Gianluca Fiori}
\affiliation{Dipartimento di Ingegneria dell' Informazione - Università di Pisa, I-56122, Pisa, Italy}
\email{*gianluca.fiori@unipi.it}
\date{\today}

\begin{document}

\begin{abstract} 
Adapting electronics to perfectly conform to non-planar and rough surfaces, such as human skin, is a challenging task, which could open up new applications in fields of high economic and scientific interest, ranging from health to robotics, human-machine interface and Internet of Things. The key to success lies in defining a technology that can lead to ultra-thin devices, exploiting ultimately thin  materials, with high mechanical flexibility and excellent electrical properties. Here, we report a hybrid approach for the development of high-performance, ultra-thin and conformable electronic devices, based on the integration of semiconducting transition metal dichalcogenides, i.e., \ce{MoS_{2}}, with organic gate dielectric material, i.e., polyvinyl formal (PVF) combined with ink-jet printed PEDOT:PSS electrodes. Through this novel approach, transistors and simple digital and analogue circuits are fabricated by a sequential stacking of ultrathin (nanometer) layers on a few microns thick polyimide substrate, which guarantees the high flexibility mandatory for the targeted applications.
\end{abstract}
%
%
%
%
\maketitle
\newpage
%
%
%
%
%

 The development of electronic circuits capable of bending and conforming to non-planar and irregular surfaces, such as human skin, is becoming essential in several applications, ranging from Internet of Things (IoT) to e-textile architectures, wearable electronics and healthcare \cite{Baran_2020, Huang_2019, Wang_2020, Bonnassieux_2021, Liu_2020, Zhang_2024}.  
 This radical change can only be made possible by conformal field-effect transistors (FETs), which in turn can become a reality through the selection of suitable materials with excellent electrical and mechanical properties, coupled with the development of frontier technologies for device fabrication beyond standard integrated circuit processes on rigid substrates.

Mechanical flexibility and conformability of materials depend not only on their intrinsic properties, i.e., bending stiffness \cite{Zhao_2020}, but also on the definition of novel methods of material film fabrication. As stiffness scales with the cube of material thickness, the possibility of employing the thinnest possible materials represents a breakthrough in conformable applications. The requirement for reduced thickness has to also match the good electrical properties of the materials as insulator (or dielectric), semiconductor, and conductors, which are the main ingredients for FET devices.

Carbon-based materials, especially organic polymers, are the current standard for flexible electronic technologies, thanks to their intrinsic mechanical flexibility and the availability of dielectric, conductive, and semiconducting organic compounds \cite{Wang_2019, Lixia_2023}. In addition, they can usually be processed using low temperature, large area and cost-effective methods, e.g., solution based, making them suitable for a wide range of applications \cite{Ling_2018,Liu_2023}. However, there are limitations to the use of these materials for the fabrication of high-performance flexible FETs and circuits, mainly related to the properties of organic semiconductors (OSCs).
The latter usually exhibit poor operational stability in ambient conditions over time \cite{Griggs_2021, Chen_2022} and mobility values below $100~$cm$^{2}$V$^{-1}$s$^{-1}$ \cite{Liu_2022, Yuvaraja_2020, Mirshojaeian_2021}.
Two-dimensional materials (2DMs), instead, such as transition metal dichalcogenides (TMDCs), with their wide range of electronic properties, from insulating to metallic or semiconducting \cite{Wang_2022, Akinwande_2014}, are the thinnest materials yet synthesized, consisting of layers just a few atoms thick, and can be easily transferred on flexible substrates. In particular, semi-conducting 2DMs, such as \ce{MoS_{2}}, exhibit exceptional electrical properties, with extremely high mobility values \cite{Pucher_2023, Alam_2020}, eventually exceeding $100~$cm$^{2}$V$^{-1}$s$^{-1}$ on FETs fabricated with standard lithographic processes \cite{Xinran_2021}, and significantly reduced stiffness (of the order of $10^{-10}$~N/m), allowing large-scale integration capabilities, with a number of integrated transistors up to ${10}^{3}-{10}^{4}~$cm$^{-2}$ \cite{Marega_2023}.

Following the FET stack, a good and as thin as possible dielectric is also required. Organic polymers are good candidates \cite{Nketia_2018} due to their mechanical properties and solution processability, which can reduce the thermal budget of the process and the fabrication costs, compared to vacuum deposition (e.g., CVD, ALD and sputtering) or oxidation techniques. However, many of these polymers have a low relative dielectric constant \cite{WangY_2019}, which limits their use in low-voltage applications. In general, they tend to exhibit electrical losses \cite{Jinhua_2018}, forcing an increase in film thickness up to the micron range, and therefore operating voltages from tens to hundreds of volts: this limits their use in portable and wearable electronics, where voltages smaller than $5~$V are required \cite{Conti_2023}. However, among the various organic polymers, poly(vinyl formal) (PVF) is a solution-processable polymer with great potential. Nanometer-scale PVF films have shown excellent insulating and mechanical properties, as well as the ability to conform to irregular and dynamic surfaces \cite{Viola_2021,Barsotti_2018, Park_2023}.

The missing element in this ambitious design is the development of a process that enables the integration of the thinnest and highest-quality materials into an ultrathin stack, completed with conductive electrodes, with the extraordinary ability to be shaped by the final application surface.

Here, we report a hybrid fabrication approach for the definition of ultrathin and high-performance conformable FETs and circuits. We integrate 2DMs and organic compounds on a flexible polyimide (PI) substrate using a combination of solution-based methods and high-quality material deposition techniques. 
We select a MOCVD grown monolayer of \ce{MoS_{2}} as the semiconducting 2DM and transfer it to PI films because it is the most promising in terms of electrical and mechanical properties. In particular, the field-effect carrier mobility of \ce{MoS_{2}} allows the definition of high-performance electronic devices \cite{Kim_2019, Piacentini_2023, Cun_2019}.  

Nanometre thick PVF films were chosen as the gate dielectric material to limit the overall thickness of the device while increasing the integration density and reducing the operating voltages. 

To define the source, drain and gate electrodes and interconnections, we chose inkjet printing due to its high customisation and versatility at room temperature \cite{Kassem_2023, Brunetti_2021, Wu_2020, Lo_2021, Karalis_2021, Hou_2023, Stritesky_2018}. It provides precise control over the volume of each droplet during the printing process, ensuring accuracy and consistency and minimising waste \cite{Carey_2017, Mitra_2017, Luczak_2022, Lemarchand_2022, Yan_2020}.
In addition, it is a scalable patterning method that provides a viable alternative to lithography in this application, where a low-temperature process is required to maintain the quality of the PVF layers. In terms of electrically conductive materials, we chose the water-based PEDOT:PSS ink, which guarantees the definition of features with thicknesses at the nanoscale, typically in the range of tens of nanometers. This order of thickness is comparable to that achieved with thermally evaporated materials. Moreover, it does not require any post-deposition baking or sintering steps, making it fully compatible with a low-temperature process. 
\begin{figure}[t]
\includegraphics[width=0.85\linewidth]{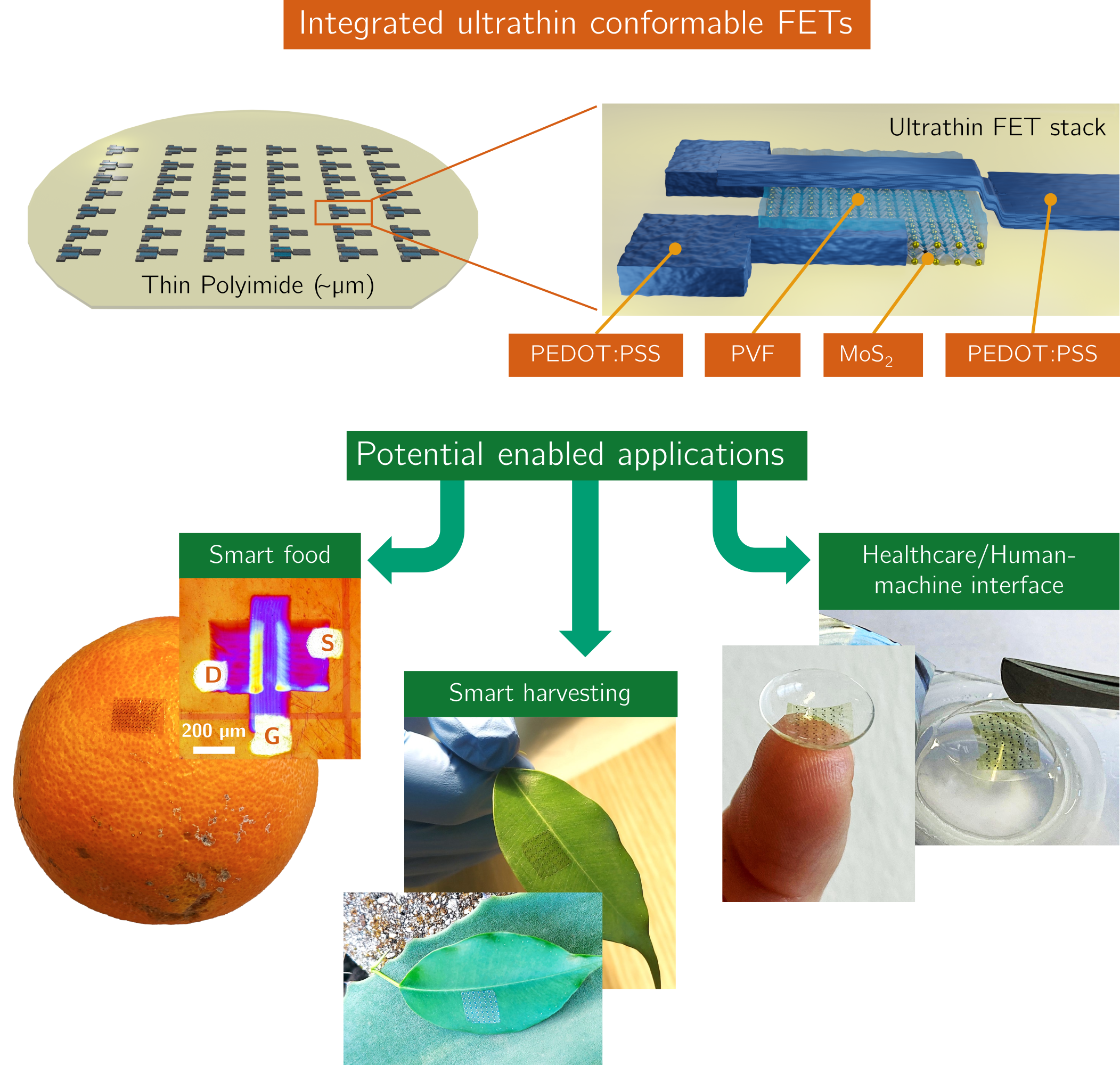}
\caption{A schematic view of an array of transistors and a sketch of a single device in the inset. At the bottom some potential applications enabled by our technology. An array of more than one hundred transistors fabricated on a polyimide flexible substrate, adhering to the surfaces of an orange (bottom left), and a leaf (in the middle), showing examples of implementation for smart agriculture. A flexible circuit (bottom right), fabricated on a polyimide substrate, conforming to an eye contact lens (dynamic surface) as an example of wearable electronic application. \textcolor{black}{In the smart food example inset, an optical microscope image of one of the devices on the orange peel is also reported.}}
\label{fig:intro_applications}
\end{figure}

This hybrid approach allowed us to define ultrathin ($<200~$nm) FETs on PI films ($3.8~\mu$m thick), with high integration density (around $100~$cm$^{-2}$) and extremely good electrical properties. 

Figure \ref{fig:intro_applications} illustrates a schematic view of a single FET, in the inset, and an array of transistors defined with our process, which have been transferred to several surfaces with a high degree of roughness, and which can open new applications, spanning from smart agriculture to wearable electronics. Based on these small devices, we have developed analog and digital circuits to showcase the potentials achievable with our technology.


\section{Fabrication of ultra-thin FETs}\label{subsec:fabrication_transistors}
\begin{figure}
\centering
\includegraphics[width=0.85\linewidth]{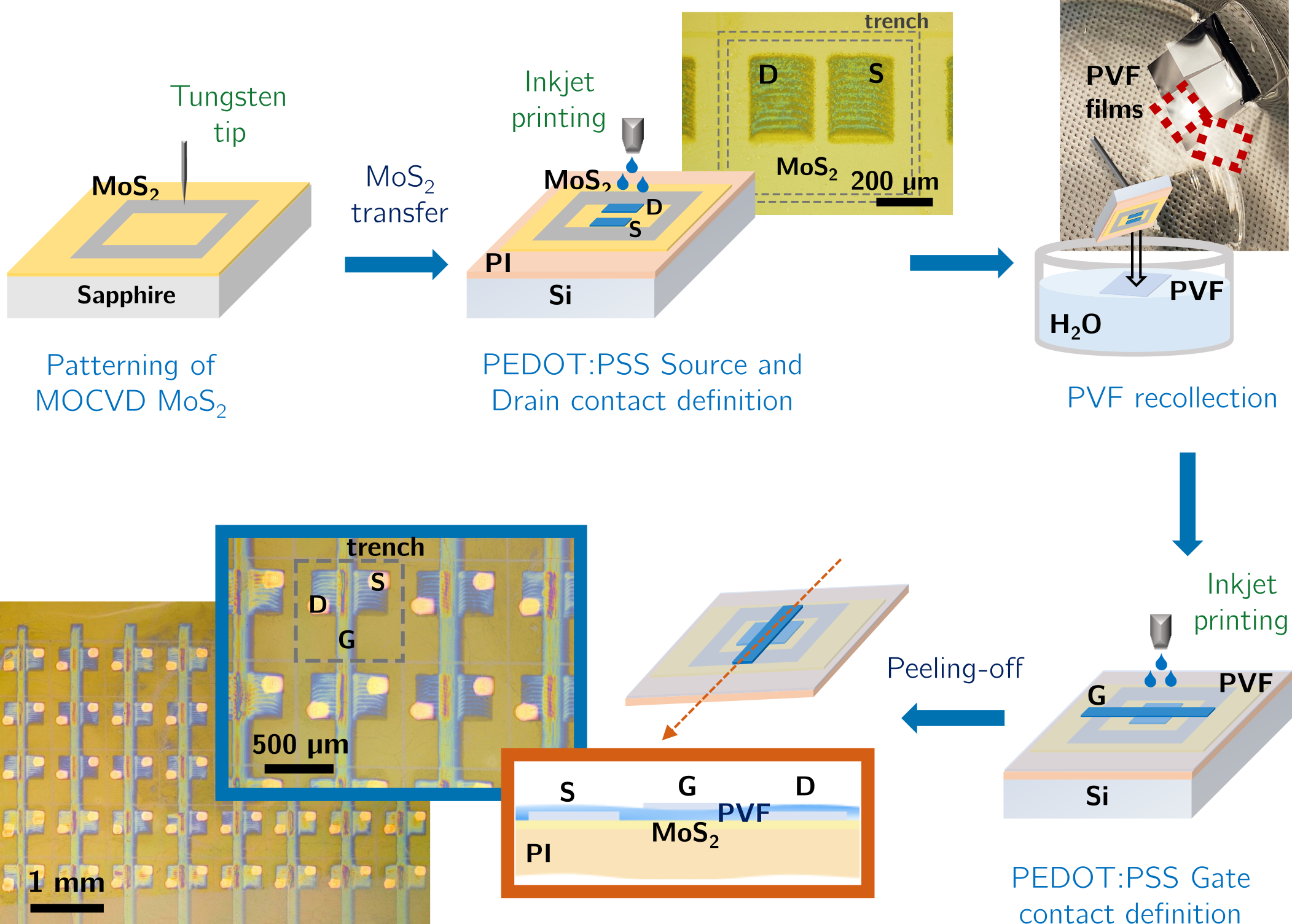}
\caption{Schematic representation of the main fabrication process steps: from the semiconductor film patterning on sapphire, to the transfer of the flexible PI substrate with the whole devices stack on top. A sketch of the final cross section
is shown in the inset. In the optical micrographs, a matrix of ultra-thin FETs fabricated on flexible PI substrate is shown.}
\label{fig:fabrication_process_steps}
\end{figure}

Ultra-thin FETs are fabricated following a bottom-up solution-based strategy, combining inkjet printing and advanced material deposition techniques, on a flexible substrate. 
Figure \ref{fig:fabrication_process_steps} illustrates the main steps of the fabrication process. 
A monolayer film of \ce{MoS_{2}}, grown through MOCVD \cite{Cun_2019}, is mechanically patterned on its native sapphire (\ce{Al_{2}O_{3}}) substrate. This patterning is achieved by precisely controlling the micrometer-scale movements of a metal tip using a custom scribing system developed in-house \cite{Sargeni_2024}. By placing the tip in contact with the target surface, it can selectively remove the 2D material in localized areas, \textcolor{black}{defining trenches with a resolution in the order of a micrometer.}
The patterned \ce{MoS_{2}} film is then transferred, using a thermal release tape, onto a few microns thick polyimide (PI) substrate, previously deposited on top of a silicon \ce{Si} wafer. 
During the transfer process, a sacrificial layer of poly(methyl methacrylate) (PMMA) is spun onto the \ce{MoS_{2}} film to provide mechanical stability and facilitate processing. After the transfer, the PMMA layer is removed, leaving a matrix of isolated \ce{MoS_{2}} areas on PI. 

Each FET is defined within an isolated semiconductor region to reduce the occurrence of high dispersion current phenomena when multiple devices are biased simultaneously (similar to a shallow trench isolation). 
On top of \ce{MoS_{2}}, transistors source and drain electrodes are printed with a water-based poly(3,4-ethylenedioxythiophene) polystyrene sulfonate (PEDOT:PSS) conductive ink, defining the region of the transistor channel with typical dimensions of $W\times L=400\times 70~\mu$m$^{2}$, where $L$ and $W$ are its length and width, respectively. 
Two $25~$nm thick PVF films, delaminated from a silicon wafer carrier and suspended in water, are collected directly with the PI substrate, with the FETs areas on top, following the procedure previously reported by \cite{Viola_2021}. 
Finally, a top-gate electrode is printed on top of the PVF using the same PEDOT:PSS conductive ink, aligned to the bottom channel areas. 
Based on the morphological analysis conducted using an Atomic Force Microscope (AFM) in the FET region, as detailed in Section 6 %
(Figure S6) %
of the Supporting Information, a comprehensive thickness assessment was achieved. This analysis revealed a total thickness of $120~$nm for the drain stack and $90~$nm for the gate stack, which is a remarkable achievement for a non-lithography-based process for conformable electronics.
The optical micrographs in Figure \ref{fig:fabrication_process_steps} show a single ultrathin FET and a dense matrix of FETs on the PI substrate fabricated through this process. \textcolor{black}{It is worth mentioning, that in the layout of the devices, the gate contact is always covering the whole channel area and protruding outside the isolated \ce{MoS_{2}} region, in order to avoid leakage currents outside the channel region and grant full control over the channel current.}
Further details of the process and the materials used can be found in Materials and Methods section of the Supporting Information.
For circuit fabrication, transistors can be interconnected by defining inkjet-printed gold vias through the insulating layer. 
Finally, the full stack can be delicately peeled from the initial substrate and transferred to different surfaces because of their exceptional conformability.

%
%
\section{Electrical characterization of conformable FETs}\label{sec:characterization_transistors}
\begin{figure}
\centering
\includegraphics[width=0.80\linewidth]{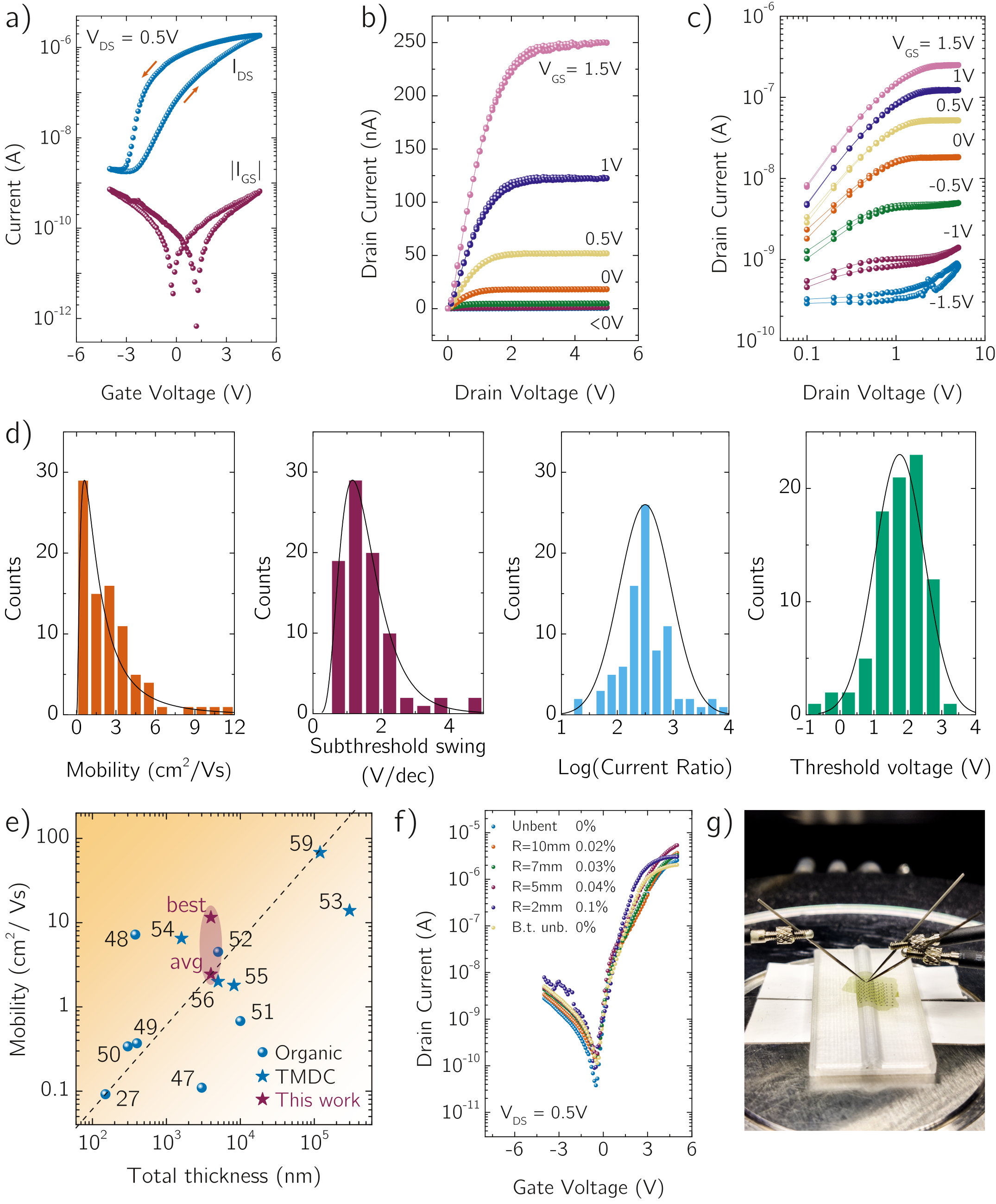}
\caption{a) Typical transfer characteristic curve of our FETs, measured for a $V_{DS}$ of $0.5~$V (scan-rate of $100~$mV/s). b) Output characteristic curves of our FETs, measured for diverse $V_{GS}$ values, and log-log curves of the output characteristics (c). d) Histograms showing mobility, threshold voltage, subthreshold swing, and $I_{max}$ over $I_{min}$ current ratio distributions, for an array of 85 transistors on the same chip. e) Comparison between the field-effect mobility related to \textcolor{black}{the thickness} of our FETs (best and average value) and other FETs on flexible substrates, based on OSCs \cite{Mirshojaeian_2023, Viola_2021, Ren_2018, Fukuda_2016, Nawrocki_2016, Zhao_2021, Mun_2023} and on TMDCs \cite{Zhao_2016, Hoang_2023, Reato_2022, Gong_2016, TANG_2023}, corresponding to blue spheres and stars, respectively. Entries have been chosen among the ones reported in Table S3 of Section 10 of the Supporting Information. \textcolor{black}{The dashed line highlights the position of our work with respect to the best entries, which are the ones located in the top left corner of the plot.} f) Transfer characteristics for diverse bending radii and $V_{DS}$ of $0.5~$V. g) Picture of the measurement setup for the bending characterization.}
\label{fig:transistors_characterization}
\end{figure}

The typical transfer and output characteristic curves, reported in Figure \ref{fig:transistors_characterization}a and Figure \ref{fig:transistors_characterization}b, demonstrate the low operation voltage range ($<5~$V) of our FETs. Moreover, the ohmic response in the low drain voltage region (Figure \ref{fig:transistors_characterization} c) suggests good electrical contact between PEDOT:PSS electrodes and \ce{MoS_{2}}.
The curves also show that the gate leakage current is negligible compared to the drain/source currents, proving the excellent quality of PVF as a gate dielectric material. To be noted that the transfer characteristic gives evidence of the typical hysteresis of 2DM based FETs, which strongly depends on the presence of charge traps at the interfaces and causes a shift in the threshold voltage from forward to backward sweep. The phenomenon is reduced in the output characteristics, which are a function of the drain-source voltage.   

As shown in Figure \ref{fig:transistors_characterization}d, the performance metrics for an array of 85 working transistors on the same chip were evaluated, showing an average threshold voltage ($V_{TH}$) of $1.76~$V, a current ratio $I_{max}/I_{min}$ typically ranging from $10^{2}$ to $10^{3}$, calculated according to the procedure reported by Cheng et al. \cite{Cheng_2022}, a subthreshold swing of $1.58~$V/dec and a mobility of $2.44~$cm$^{2}$V$^{-1}$s$^{-1}$, for devices with a channel width and length of $400~\mu$m and $70~\mu$m, respectively. These parameters were extracted following the procedure reported in Section 7 
of the Supporting Information, and the related descriptive statistics are summarized in Table S2 
of the same Section.

The field-effect mobility ($\mu_{FE}$), a crucial factor in assessing the electric performance of a FET, was evaluated according to the bias condition of the devices, such as linear regime or saturation regime, employing the expressions derived for an ideal long-channel MOSFET for $V_{GS}>V_{TH}$:
\begin{equation*}\label{eq:mobility}
\mu_{FE} = \genfrac{\{}{.}{0pt}{}{\frac{L}{W}\frac{1}{C_i}\frac{1}{V_{DS}}\frac{\partial{I_D}}{\partial{V_{GS}}}, \textrm V_{DS}<V_{GS}-V_{TH}}{2\frac{L}{W}\frac{1}{C_i}(\frac{\partial{\sqrt{I_D}}}{\partial{V_{GS}}})^2, \textrm V_{DS}>V_{GS}-V_{TH}}
\end{equation*}
where $C_{i}$ is the insulator film capacitance per unit area, $V_{GS}$ is the gate to source voltage and $V_{DS}$ is the drain to source voltage. 

To guarantee a precise mobility estimation, the capacitance value was calculated on a PVF-based parallel plate capacitor structure, fabricated on PI.
Following the classical expression:
\begin{equation*}\label{eq:capacitance}
C_{i} = \frac{\epsilon_{r}\epsilon_{0}}{t}
\end{equation*}
where $\epsilon_{r}$ is the relative permittivity of the insulator, $\epsilon_{0}$ is the permittivity of free space and $t$ is the insulator film thickness.

By measuring the capacitance per unit area, using an insulator thickness of $50~$nm common to all fabricated devices, it was found that the average relative permittivity value was 3.8. This result is in agreement with other characterizations of the material reported in previous studies \cite{Barsotti_2018,Viola_2021}.
More details and data about capacitance measurements can be found in Section 8
of the Supporting Information. 
Taking into account this value of capacitance per unit area, an average field-effect mobility of $2.44~$cm$^{2}$V$^{-1}$s$^{-1}$ was estimated. \textcolor{black}{The good mobility and the small operating voltage} achieved with our devices demonstrate the robust competitiveness in terms of electrical performances, which can lead to a low power consumption, crucial for portable applications. 

\textcolor{black}{For conformable electronic applications, the total device thickness ($t_h$ including the gate stack, the semiconductor and the substrate) is a fundamental parameter. As a consequence, we choose mobility and the total thickness as metrics to compare our work with other results found in the literature.} In Figure \ref{fig:transistors_characterization}e, the best and average field-effect mobility ($\mu_{FE}$), \textcolor{black}{expressed as a function of the total thickness of our FETs}, is compared with the values reported by other groups for transistors defined on flexible substrates. Only devices with a FET stack thickness below $1~\mu$m have been considered.  
Most of the reported values are referenced to organic semiconductor-based FETs, indicated with blue spheres, as representative of the current standard technology for flexible electronics. Entries for TMDC-based flexible FETs, indicated with blue stars, are also included. Following the color gradient, the top left area indicates the small thickness and high-mobility ($>1~$cm$^{2}$V$^{-1}$s$^{-1}$) region, where our work is located. 
\textcolor{black}{The dashed line is described by the expression $\mu = k \cdot t_h$, where $\mu$, $t_h$ and $k$ represent the mobility, the total thickness and the ratio between our mobility and thickness values, respectively. In our case, the value of $k$ is $6.1 \times 10^3$~cmV$^{-2}$s$^{-1}$. This line represents a guide to the eye and highlight the position of our work among the best entries, and it does not indicate a physical relationship between the two metrics. It does not indicate a physical relationship between the two metrics.} The entries above this line, following the color gradient, denote instances where the ratio $\mu$ to \textcolor{black}{$t_h$} exceeds our own. 
It is worth noting that, differently to our case, these instances involve lithographic-based fabrication processes and vacuum-based advanced deposition techniques, which inherently imply higher fabrication costs and higher thermal budgets, the latter preventing device fabrication on top of substrates of interest for recyclable and responsible electronics as the paper.
\textcolor{black}{However, the obtained average mobility value, which is lower than the expected one for this \ce{MoS_{2}} \cite{Cun_2019}, has been further investigated to find a possible weak link in our process chain.} 

\textcolor{black}{A comparative analysis on the performance of FETs between our top-gate structure and an alternative bottom-gate structure, fabricated on a highly doped \ce{Si}/\ce{SiO_{2}} wafer, based both on transferred MOCVD \ce{MoS_{2}} as semiconductor, has been reported in the Section 7
of the Supporting Information. 
Moreover, we compared two different inks, PEDOT:PSS and \ce{Ag}, as conductive materials for the inkjet-printed top electrodes (source/drain) in the back-gate configuration. This analysis shows that the average mobility values for the alternative configurations are of the same order of magnitude as those obtained with top-gate devices fabricated on PI with PVF as the gate dielectric.} 

\textcolor{black}{These results seem to imply that the measured mobility values are not significantly impacted by the contact or dielectric material choice.}

To demonstrate the pliability and conformability of our devices, we performed an electromechanical characterization. This involved assessing their electrical response when subjected to static bending conditions with various curvature radii, thereby demonstrating their ability to maintain functionality even when conformed to different shapes. Figure \ref{fig:transistors_characterization}f shows multiple transfer characteristics for bending radii \textcolor{black}{from $10~$mm to $2~$mm, compared with the unbent conditions}, confirming that the electrical response of the devices remains unaffected by the bending condition, as no significant changes are observed in the drain and gate currents.  
In Figure \ref{fig:transistors_characterization}g, a picture of our setup for electromechanical characterization is reported.
\textcolor{black}{In section 9 of the supporting information, we have also reported the results of an electrical characterization of the \ce{MoS_{2}}-PEDOT:PSS-PVF transistors transferred on an orange peel, in order to prove their functionality while adhering to irregular, rough and bent surfaces.}
Finally, devices must be able to function optimally even when conformed to nonplanar dynamic surfaces, such as human skin, which may subject them to repeated bending cycles. 
Hence, we conducted an investigation into the longevity of our devices, observing negligible alterations in their electrical behavior even after subjecting them to numerous bending cycles (up to 500), underscoring their robustness and durability. Results and more details can be found in Section 9 
of the Supporting Information.
Similarly, a parallel study was conducted on PEDOT:PSS-PVF parallel plate capacitors, giving results consistent with the ones derived from the FETs analysis. This parallel investigation reinforces the conclusions on the performance and durability of our devices and confirms the robustness of our technology across different device architectures.
%
%
\section{Conformable electronic circuits}\label{subsec:characterization_circuits}
\begin{figure}
\centering
\includegraphics[width=0.85\linewidth]{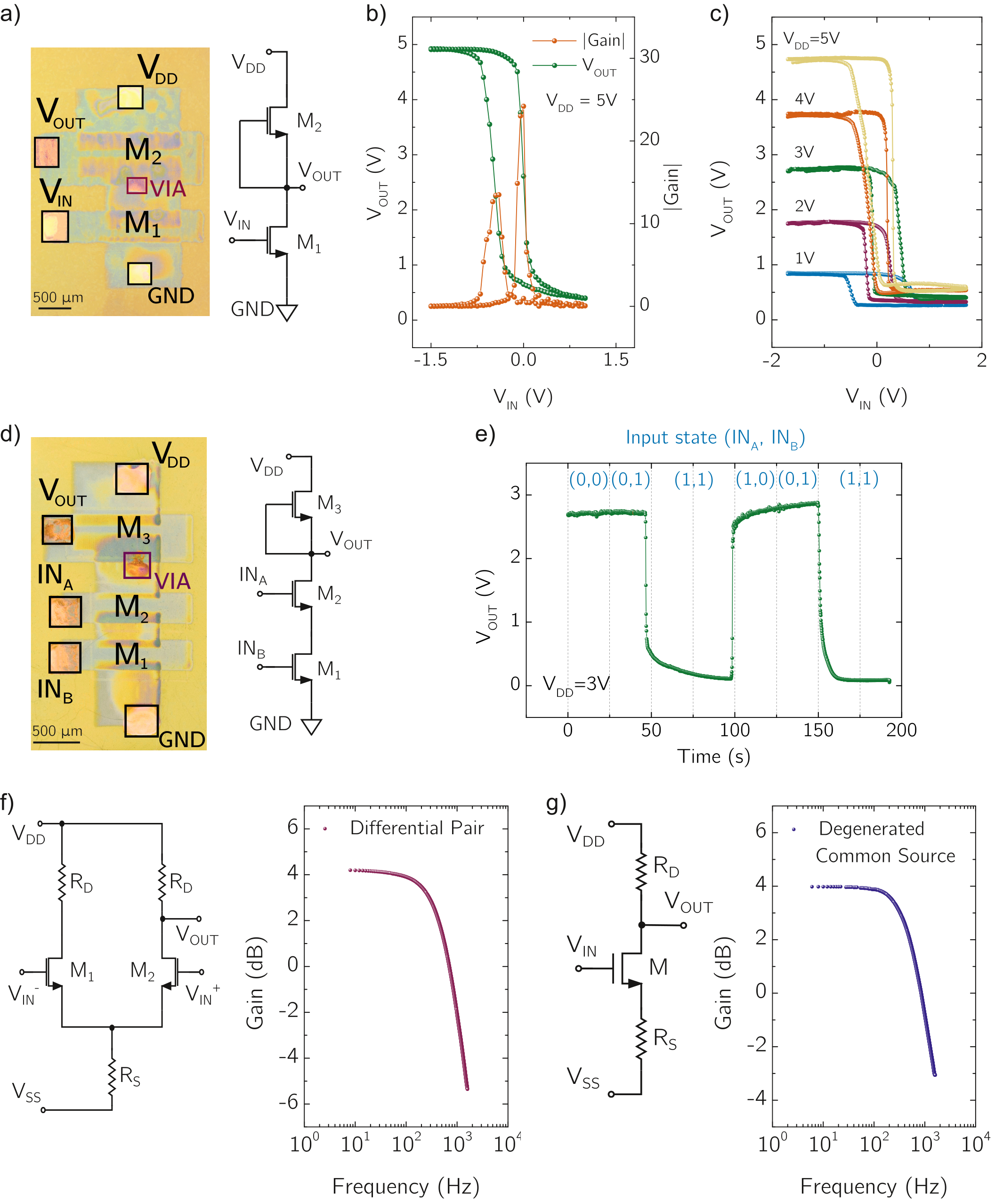}
\caption{a) Optical micrograph and electrical schematic of a depletion-load inverter. b) Transfer characteristic (left axis) and voltage gain modulus (right axis) of a depletion-load inverter gate as a function of the input voltage, for a supply voltage of $5~$V, and for diverse values of supply voltage, down to $1~$V (c) d) Optical micrograph and electrical schematic of a depletion-load NAND gate. e) Output voltage of the NAND gate as a function of the input sequence (${V}_{{IN}_{1}}$,${V}_{{IN}_{2}}$), with a supply voltage of $3~$V. f) Electric schematic of a differential pair (left) and its frequency response (right). g) Electric schematic of a degenerated common source stage (left) and its frequency response (right).}
\label{fig:circuits_characterization}
\end{figure}

Several circuits have been fabricated using the proposed technology and the previously described FETs as elementary building blocks. Figure \ref{fig:circuits_characterization}a shows the optical micrograph and the electrical schematic of an inverter logic gate composed of two transistors, $M1$ and $M2$, following the depletion-load nMOS-like logic, where the transistor $M2$ acts as a pull-up resistor.
The aspect ratio ($W/L$) is about $17$ for $M1$ and $19$ for $M2$.

Figures \ref{fig:circuits_characterization}b and c show the input-output characteristics measured for a single value of the supply voltage ($V_{DD}$) of $5~$V and for several values ($5$~V down to $1~$V), respectively.
The gain ($G$), defined as the slope of the transfer curve ($dV_{OUT}$/$dV_{IN}$), where $V_{IN}$ and $V_{OUT}$ are the input and output voltages, is also shown (right axis).  The inverter has a high gain value of $25$ when the circuit is biased with a voltage of $5$~V, and it maintains almost full output swing even when biased with smaller values of $V_{DD}$, down to $1~$V. The inversion voltage decreases as $V_{DD}$ is reduced, along with the gain.
Another interesting digital circuit is the NAND gate, which is essential in combinational logic as it can be used to implement all the other logic functions. Hence, defining a conformable NAND enables all boolean operations for any conformable application.
Figure \ref{fig:circuits_characterization}d shows the schematic and optical micrograph of a depletion-load nMOS-like NAND gate, where transistors $M1$ and $M2$ act as a pull-down network, and $M3$ as pull-up resistor. 
The aspect ratio is about $7.5$ for $M1$ and $M2$, and $25$ for $M3$. Figure \ref{fig:circuits_characterization}e shows the circuit output voltage as a function of the input sequence ($V_{{IN}_{1}}$,$V_{{IN}_{2}}$). The input high-logic value corresponds to $3~$V, while the low level to $0~$V. The power supply $V_{DD}$ is set to $3~$V. Accordingly to the truth table of a NAND gate, the output voltage is in low state ($0$) only when both input signals, represented by $V_{{IN}_{1}}$ and $V_{{IN}_{2}}$, are in high state ($1,1$). Otherwise, output voltage results to be in high state ($1$). The output voltage plot confirms that this condition is satisfied by our device. Moreover, the output state transitions are steep and the output swing is almost full. 
To demonstrate the versatility of our technology, we have also defined analogue circuits. The schematics of a fabricated differential pair and a degenerated common-source stage \cite{raz_2021} are presented in Figure \ref{fig:circuits_characterization}f and g, respectively. Transistors $M_1$, $M_B$ and $M$ are all fabricated with an aspect ratio of about $7$, the supply voltages applied are $\pm 10~$V, and the values of resistances $R_{D}$ and $R_{S}$ are $8.2~$M$\Omega$ and $1~$M$\Omega$, respectively. In addition, the corresponding frequency responses, illustrating gain as a function of frequency, are provided for both circuits. This characterization demonstrates the capability of our technology to produce analog circuits with desirable performance.
%

To summarize, in this study, we have successfully defined ultra-thin and highly conformable field-effect transistors (FETs) using a low-cost and low-temperature fabrication process, which combines high-quality MOCVD-grown monolayer \ce{MoS_{2}} with solution-processable organic materials such as PI, PVF and PEDOT:PSS. 
Our approach involves the sequential stacking of nanometer-scale layers of flexible materials, resulting in a gate-stack structure with a total thickness of $90~$nm. Notably, this achievement surpasses results obtained by processes that do not include lithographic steps. The fabricated FETs exhibited exceptional performance characteristics, facilitating their integration into more complex electronic circuits for both digital (e.g., depletion load inverter and NAND gates) and analog (e.g., differential pairs and degenerated common source amplifiers) applications. 
Furthermore, our investigation demonstrated that these devices maintain satisfactory operation under bending stress and repeated bending cycles, with minimal impact on their characteristics. This resilience is crucial for applications in conformable electronics, affirming the validity of our approach.



%
\subsection{Acknowledgements}
Authors gratefully acknowledge the ERC CoG PEP2D (Contract No. 770047), the Italian Ministry of Education and Research (MIUR) in the framework of the FoReLab project (Departments of Excellence) and the Piano Nazionale di Ricerca e Resilienza (PNRR). 
This work was financially supported by the European Union’s Horizon 2020 research and innovation program under grant agreement No 964735 (EXTREME-IR), the Swiss National Science Foundation (grant 170748) and the CCMX Materials Challenge grant “Large area growth of 2D materials for device integration."
%
%
%
\newpage

%

%

%

%
%
\section{Ultra-thin transistors and circuits for conformable electronics -- Supporting Information} 


\section{Materials and Methods}\label{sec:methods}
\subsection*{Ultra-thin FETs fabrication}\label{subsec:FETs_fabrication}
Field-effect transistors were fabricated with a top-gate/top-contact configuration on \ce{MoS_{2}} films transferred onto polyimide (PI) substrates.
PI substrates were defined starting from solution (PI2611, purchased from HD Microsystems) deposited on top of silicon chips, with a film thickness of $3.8~\mu$m, following the procedure described in Section "Polyimide films deposition and morphological characterization". 
MOCVD-grown \ce{MoS_{2}} films were grown as reported by Cun at al.\cite{Cun_2019}.

\paragraph*{Molybdenum disulfide patterning}\label{par:MoS2_patterning}
Isolated rectangular regions of \ce{MoS_{2}} films on the native sapphire substrate were defined by mechanically scratching the surface with a high-precision materials printer equipped with a scratching lithography tool. 
This step ensured insulation between neighbouring devices on the same substrate, reducing leakage current during biasing. Then, the \ce{MoS_{2}} films were transferred on the PI substrates following the procedure described in Section "Molybdenum disulfide films transfer". Raman analysis of \ce{MoS_{2}} before and after the transfer process has been performed to prove the high quality of this semiconducting layer. The complete analysis has been reported in Section "Raman analysis"
(Figure \ref{fig:raman_02}). 

\paragraph*{Inkjet printing}\label{par:inkjet_printing}
Inkjet printing was used for the definition of source, drain, gate electrodes and interconnections. A Fujifilm Dimatix DMP2850 equipped with $2.4~$pL Samba nozzle cartridges was used to print patterns with a PEDOT:PSS conductive ink (RD CleviosTM P Jet X N, purchased from Heraeus). The conductivity of PEDOT:PSS and the dispersion viscosity were enhanced by incorporating anhydrous ethylene glycol ($99.8\%$ by Sigma Aldrich) at a $5\%~wt.$ concentration. Then, a non-ionic polyoxyethylene surfactant solution (TritonTM X-100 by Sigma Aldrich) was added to improve the wettability of the ink at a $1\%~wt.$ concentration.
All electrodes and patterns were printed in one layer with a drop spacing of $25~\mu$m, with the printer platen temperature heated at $40~^{\circ}$C. No annealing or post-treatment processes were performed after any of the printing steps. 
\paragraph*{Polyvinyl formal deposition}\label{par:PVF_deposition}
Once defined the bottom source and drain electrodes on top of the \ce{MoS_{2}} areas, the PVF nanosheets were recollected following the procedure described in Section "Polyvinyl formal films deposition and transfer". The procedure was repeated twice to build a double layer stack of PVF nanosheets, with a final thickness of $50~$nm, with the purpose of improving the dielectric reliability, i.e., reducing the leakage current of the final transistors. Raman analysis of the structure with PI/\ce{MoS_{2}}/PVF has also been performed and reported in Section "Raman analysis"
(Figure \ref{fig:raman_01}).
\subsection*{Circuits Fabrication}\label{subsec:circuits_fabrication}
For interconnecting transistors, PEDOT:PSS and a commercial water-based gold ink (DryCure Au-J, purchased from C-INK Co., Ltd.) were used to create vias through the insulating layer. These vias enabled the connection of top gate electrodes to bottom source and drain electrodes.
To test the device properties under different bending conditions, the PI films with the integrated devices on top were peeled from the initial rigid substrates and transferred on cylindrical surfaces with different curvature radii.
\subsection*{Measurements}\label{subsec:measurements}
All electrical measurements were conducted under ambient conditions. DC characterization of transistors and circuits was performed using a Keithley 4200 SCS parameter analyzer, multiple Keithley 2450 source meter units, a Tektronix MSO2014B oscilloscope, a HP 33120A function/arbitrary waveform generator and an ONO SOKKY CF-9400 FFT analyzer. Capacitance measurements were carried out with a Keysight E4989A LCR meter.

\section{Polyimide films deposition and morphological characterization}\label{sec:PI_deposition_morphocar}
Polyimide films were deposited on \ce{SiO_{2}}/\ce{Si} substrates of approximately $1~$cm$^{2}$ area.
The chips were first cleaned in acetone and isopropyl alcohol (IPA), then treated with $15$ minutes UV/ozone (UVO) to enhance surface hydrophilicity, followed by dehydration for 15 minutes at $135~^{\circ}C$ on a hotplate. 
The polyimide solution (PI2611, purchased from HD Microsystems) was spun at $3000~$rpm for $40$ seconds on top of the cleaned chips, achieving a film thickness of $3.8~\mu$m. A two-step soft baking process ($65~^{\circ}$C for $3$ minutes, then $135~^{\circ}$C for $3$ minutes) was performed to remove excess solvent. Finally, a two-step curing treatment was performed in a quartz tube furnace in \ce{N_{2}} atmosphere ($1000~$sccm): $200~^{\circ}$C for $1$ hour, then $300~^{\circ}$C for $2$ hours, allowing polymerization and cross-linking. The curing temperature sets the maximum processing temperature that the PI film can withstand, with the glass transition occurring at $380~^{\circ}$C. A ramp rate of $2~^{\circ}$C per minute ensured low residual mechanical stress, producing high-quality films. The deposited PI films showed complete compatibility with solution-based methods and low-temperature (below the curing set point) processes.

\begin{figure}
\centering
\includegraphics[width=0.9\linewidth]{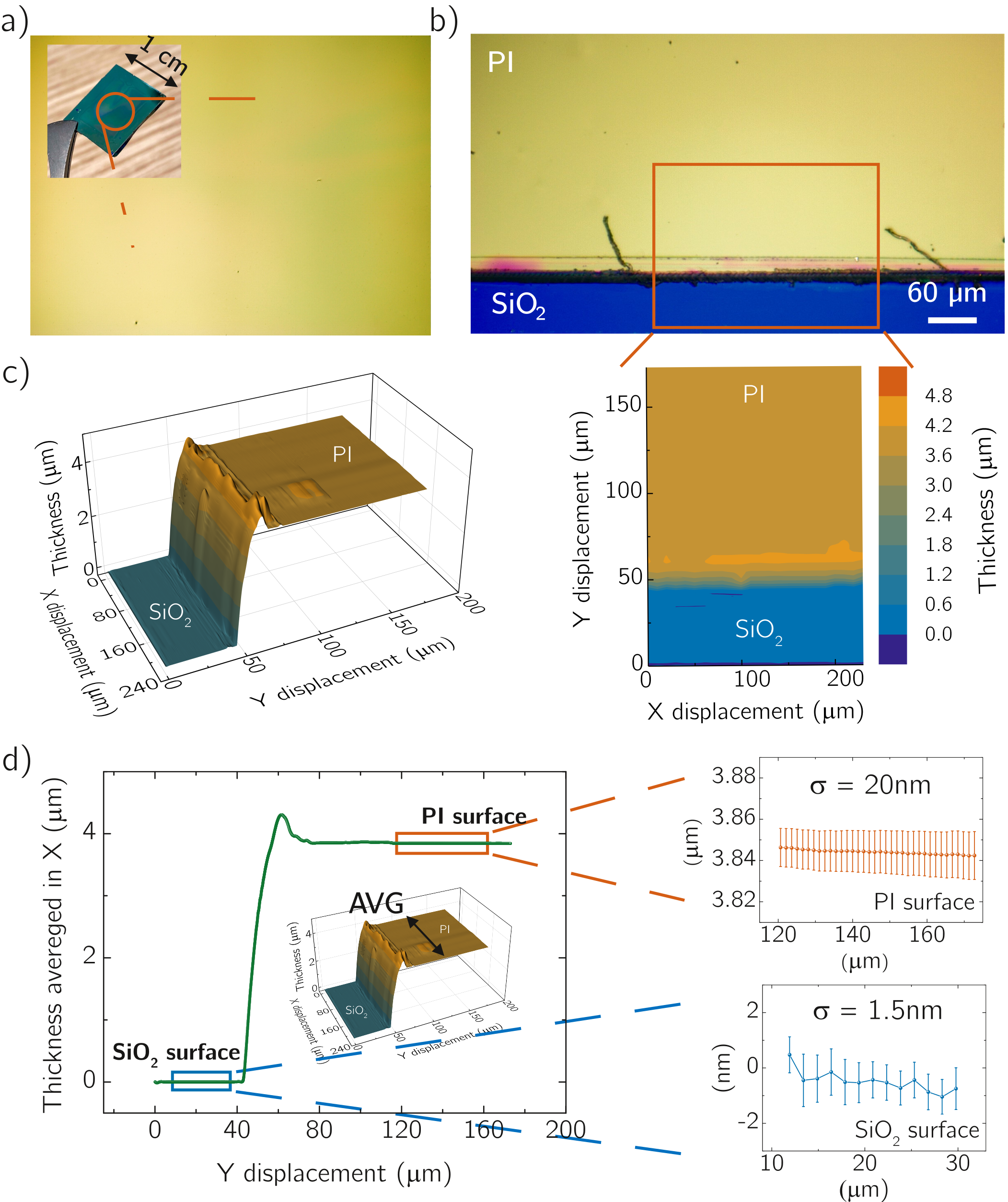}
\caption{a) Micrograph of a PI film surface in its central region. In the inset, a picture of the sample, handled with tweezers. b) Micrograph of the scratch made with a surgery blade on the PI film for the morphological analysis. c) Morphological map of a $230\times175~\mu$m$^{2}$ area as a result of the microprofilometer measurements. The thickening of the PI film at its edge is attributed to the formation of wrinkles due to the stress caused by the blade. d) PI film profile along the Y direction, derived from the map averaging in the X direction. On the right, detailed views of the regions of PI and \ce{SiO_{2}}. The error bars represent the standard deviation ($\sigma$) of the thickness.} 
\label{fig:PI_characterization}
\end{figure}

Figure \ref{fig:PI_characterization}a shows an image of a typical PI film spun over \ce{SiO_{2}}, which was used as a substrate for our process. 
A morphological analysis of a $230\times175~\mu$m$^{2}$ scratched area of the PI film, reported in Figure \ref{fig:PI_characterization}b, was performed with a microprofilometer (Bruker DektakXT) operated in the 3D Map mode. 
The analysis revealed an average thickness of $3.8~\mu$m with minimum fluctuations on the PI surface. 
Figure \ref{fig:PI_characterization}d illustrates the thickness of the PI film as it varies with the measurement direction ($Y$), obtained from the morphological map represented in Figure \ref{fig:PI_characterization}c by averaging in the transversal direction ($X$) of the profiles. The error bars depict thickness fluctuations, approximately $1.5~$nm for the \ce{SiO_{2}} surface and $20~$nm for the PI surface.

\newpage
\section{Polyvinyl formal films deposition and transfer}\label{sec:PVF_deposition_transfer}
Poly(vinyl formal) (PVF) nanometric films fabrication was performed adapting the processes previously reported by Viola et al. \cite{Viola_2021} and Baxamusa et al.\cite{Baxamusa_2014}.

For the preparation of the dielectric layer, two solutions were needed: one for the PVF layer, purchasing the starting powder from SPI Supplies as Vinylec E Polyvinyl Formal Resin, and a second for the poly(diallyldimethyl ammonium chloride) (PDAC) layer (purchased in solution, $20\%~wt.$ in deionized water, from Sigma Aldrich). 
A $0.5\%~wt.$ PDAC solution was obtained by dilution in deionized water and then stirred for $15$ minutes. A $1\%~wt.$ PVF solution was prepared dissolving the PVF powder in ethyl lactate (EL) by stirring at $650~$rpm and $50~^{\circ}$C for $3$ hours. Before use, the PVF solution was heated to $50 ^{\circ}$C and stirred, again at $650~$rpm, for $15$ minutes to prevent polymer aggregates.

A \ce{Si} wafer, used as substrate, was initially cleaned with acetone, followed by IPA, then treated with $15$ UVO to improve the wettability of its surface. Immediately after, the PDAC solution was spun on its top at $4000~$rpm for $15$ seconds, followed by $10$ seconds of baking on a hot plate at $100~^{\circ}$C, resulting in a subnanometric layer. The PVF solution was spun over the PDAC layer in a two-step process: $5$ seconds at $300~$rpm and $5$ seconds at $3000~$rpm. The PVF and PDAC layers were then baked on a hot plate at $50~^{\circ}$C for $60$ seconds to remove the excess solvent, achieving a thickness of $25~$nm. 
The PVF/PDAC layers were then cut in squares and transferred from the carrier to the receiver substrates, taking advantage of the hydrosolubility of the PDAC interlayers. The carrier wafer was slowly dipped in deionized water at an angle of approximately $45^{\circ}$, inducing the complete dissolution of the PDAC interlayer. Finally, the PVF nanosheets, floating on the water surface due to their hydrophobicity, were drawn directly with the final substrates. 

\begin{figure}
\centering
\includegraphics[width=0.9\linewidth]{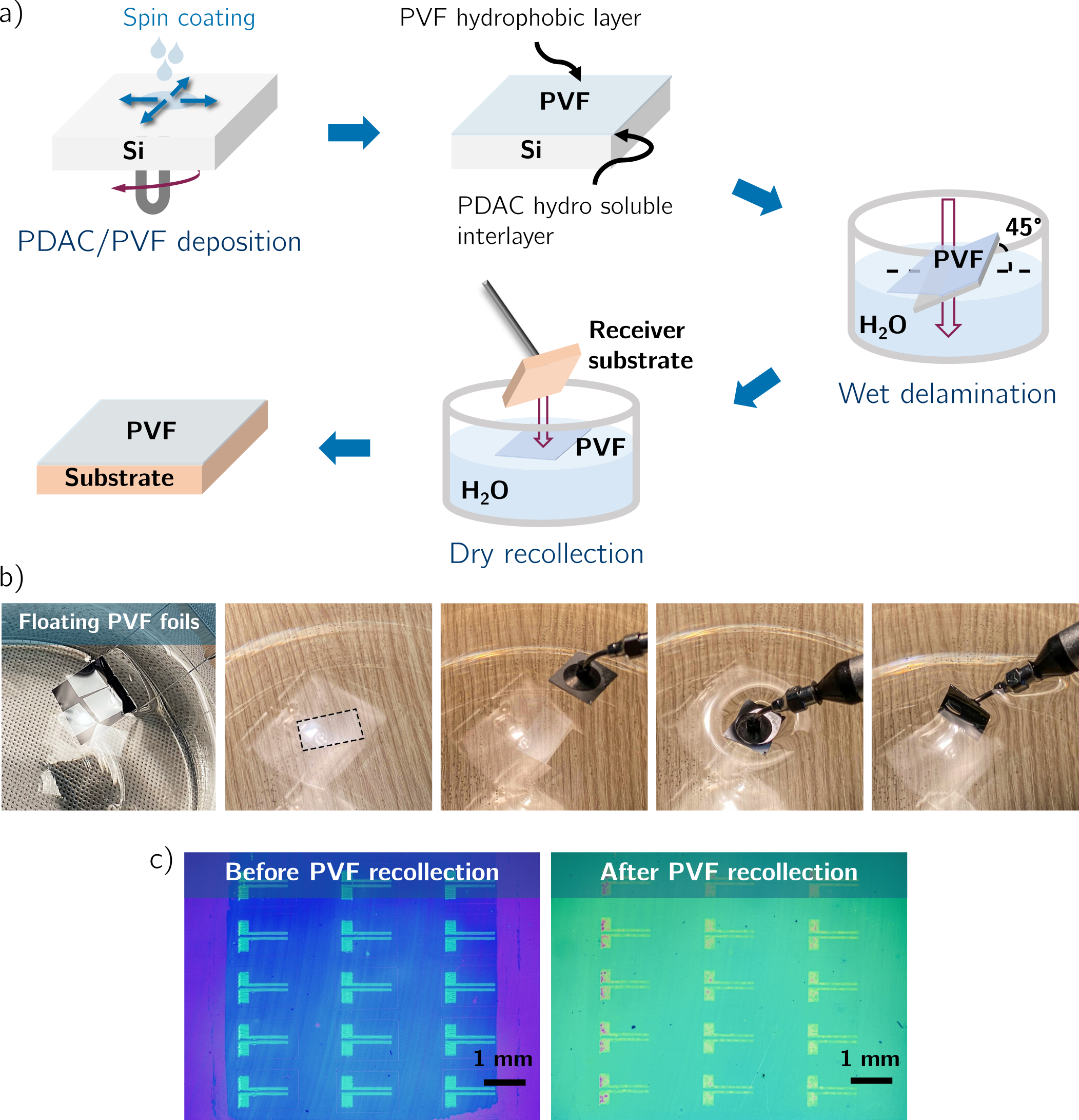}
\caption{a) Schematic representation of the PVF deposition and transferring process. b) Pictures taken during the PVF nanosheets transfer process. From the PVF sheets floating on the deionized water surface, after the delamination from their carrier \ce{Si} wafer (on the left), to the single PVF sheet (marked with a black dotted rectangle) recollection, once in close contact with the receiver substrate (on the right). c) Micrographs showing an example of a PVF film transfer on a \ce{Si/SiO_{2}} substrate with \ce{MoS_{2}} and PEDOT:PSS electrodes on top (left).}
\label{fig:PVF_process_steps}
\end{figure}

Figure \ref{fig:PVF_process_steps}a and b show the schematic representation of the process and some pictures taken during the delamination and recollection phases.

\newpage
\section{Molybdenum disulfide films transfer}\label{sec:MoS2_growth_transfer}
For the transfer process, a layer of poly(methyl methacrylate) (PMMA) was spun over \ce{MoS_{2}} grown on sapphire substrates, to provide mechanical stability and preserve the quality of the 2DM layer and to ease its detachment. A piece of commercial thermal-release tape (3195MS purchased from REVALPHA) was attached to the substrate and immersed in deionized water for $10$ minutes. The stack composed by the thermal tape with PMMA and the embedded \ce{MoS_{2}} was then peeled off from the sapphire and placed on the PI/\ce{Si} substrate on a hotplate at $50~^{\circ}$C, for $30$ minutes. The thermal tape was then released, raising the temperature to $130~^{\circ}$C.  
Finally, the sacrificial layer of PMMA was removed with a bath in hot acetone ($50~^{\circ}$C) for $1$ hour and a rinsing in IPA. 

\begin{figure}
\centering
\includegraphics[width=0.9\linewidth]{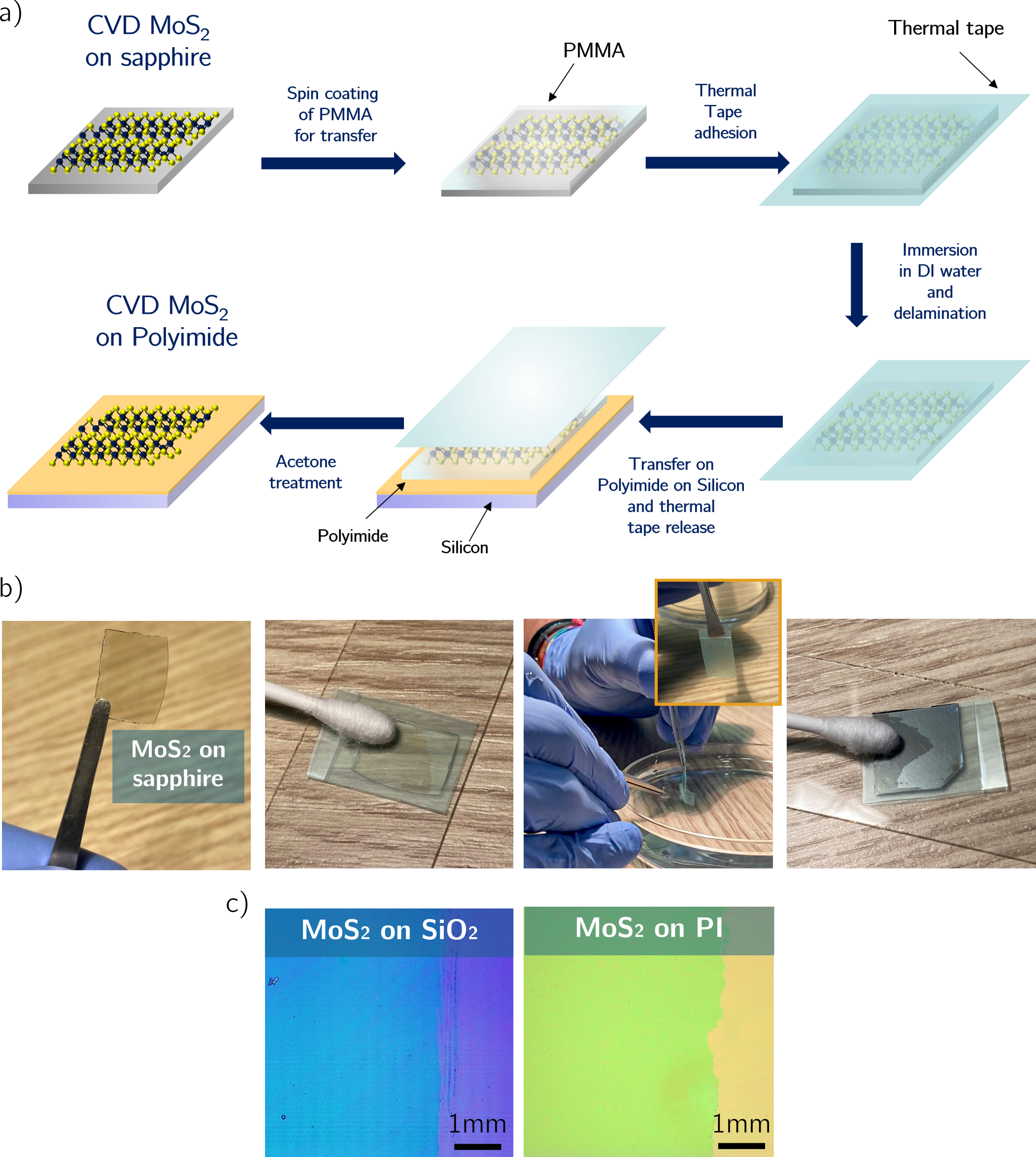}
\caption{a) Schematic representation of a \ce{MoS_{2}} film transferring process. b) Pictures taken during the process. From left to right: Sapphire substrate with \ce{MoS_{2}} on top, first thermal release tape adhesion on PMMA/\ce{MoS_{2}}/sapphire stack (a cotton swab helps to avoid formation of bubbles), peeling of the tape/PMMA/\ce{MoS_{2}} stack from sapphire, and its transfer on PI through the second adhesion (a cotton swab is used again with the same purpose). c) A \ce{MoS_{2}} film transferred on top of \ce{SiO_{2}} (left) and on PI (right) substrates.}
\label{fig:MoS2_transfer_process_steps}
\end{figure}

The schematic representation of the process together with some pictures taken during its progress are reported in Figures \ref{fig:MoS2_transfer_process_steps}a and \ref{fig:MoS2_transfer_process_steps}b, respectively. In Figure \ref{fig:MoS2_transfer_process_steps}c, the \ce{MoS_{2}} film transferred on top of \ce{SiO_{2}} (left) and of PI (right) substrates is shown.  

\newpage 
\section{Raman analysis}\label{sec:raman_analysis}
Scanning Raman spectroscopy was carried out with a Renishaw InVia system, equipped with a confocal microscope, a $532~$nm excitation laser and a $2400~$line/mm grating (spectral resolution $<1~$cm$^{-1}$). All analysis were performed with a 100X objective ($NA=0.85$), an excitation laser with a power of $500~\mu$W and an acquisition time of $3$ seconds. The Raman modes were fitted with a Lorentzian peak.

\begin{figure}
\centering
\includegraphics[width=0.9\linewidth]{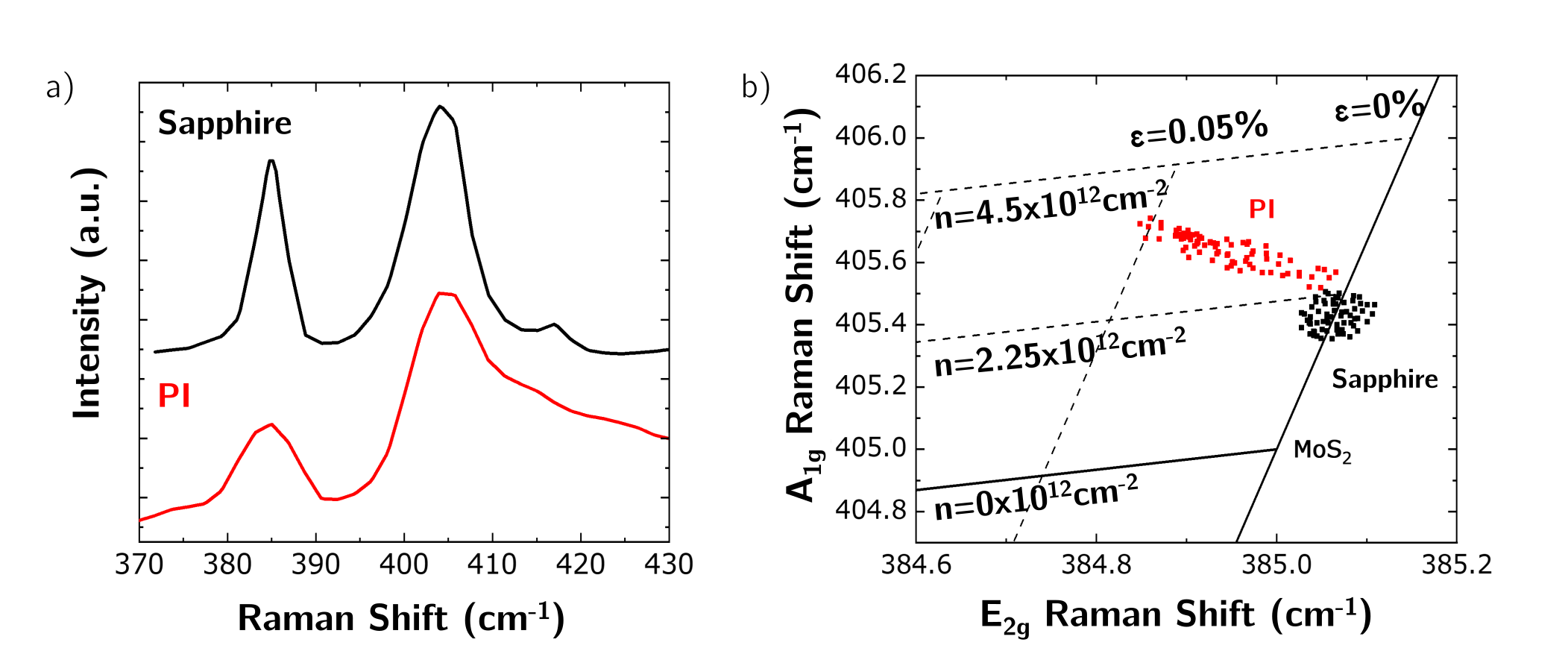}
\caption{Comparison of the Raman data of \ce{MoS_{2}} monolayer on sapphire and on PI. a) Representative Raman spectra, on sapphire (black line) and on PI (red line). b) \ce{MoS_{2}} $\epsilon-n$ correlation plot for the evaluation of the strain and carrier concentration in case of the \ce{MoS_{2}} grown on sapphire (black dots) and transferred on PI (red dots).}
\label{fig:raman_02}
\end{figure}

Figure \ref{fig:raman_02} presents the Raman characterization of the monolayer \ce{MoS_{2}}. The representative Raman spectra (Figure \ref{fig:raman_02}a), of \ce{MoS_{2}} on the sapphire growth substrate (black line) and of \ce{MoS_{2}} transferred to polyimide with PVF on top (red line), present the standard Raman modes, $E_{2g}$, at $385.1~$cm$^{-1}$ and $A_{1g}$, at $405.2~$cm$^{-1}$. The $A_{1g}$ mode corresponds to the sulfur atoms oscillating in anti-phase out-of-plane and the $E_{2g}$ mode is related to the sulfur and molybdenum atoms oscillating in anti-phase parallel to the crystal plane \cite{Lloyd_2016}.

To quantify strain and doping after the transfer process, we employed the \ce{MoS_{2}} correlation plot of the Raman shifts of modes $E_{2g}$ and $A_{1g}$, also known as the $\epsilon-n$ system \cite{Michail_2016, Michail_2018}. This method allows to disentangle and to quantify the strain and doping variations and it is normally employed for studying the effect of different growth substrates \cite{Chae_2017} or in transferred \ce{MoS_{2}} for the development of Van der Waals heterostructures \cite{Ciampalini_2022}. The full lines represent the zero strain and zero doping lines, while the dashed lines correspond to iso-strain and the iso-doping lines, calculated following the insights from previous works \cite{Lloyd_2016, Chakraborty_2012}. 

The origin of the system $\epsilon-n$ is the zero strain and charge neutrality phonon frequencies, which are set at $385~$cm$^{-1}$ for the $E_{2g}$ mode and at $405~$cm$^{-1}$ for the $A_{1g}$ mode, evaluated in the case of CVD-grown \ce{MoS_{2}} suspended monolayer membrane \cite{Lloyd_2016}. Data on the \ce{MoS_{2}} monolayer grown on the sapphire substrate present a round distribution revealing a strain-free monolayer and an average positive charge concentration of $(2.24 \pm 0.05) \cdot 10^{12}~$cm$^{-2}$. While, in the case of the \ce{MoS_{2}} transferred on polyimide, the data distribution is horizontally dispersed. Tensile strain varies between strain-free and to a maximum value of $0.05\%$. 
The charge concentration increases, showing a variation between $2.3 \cdot 10^{12}~$cm$^{-2}$ and $3.4 \cdot 10^{12}~$cm$^{-2}$. The increase of the tensile strain is probably due to the morphology of the polyimide layer, which has a larger roughness compared to sapphire. The increase in electron concentration is related to a possible charge transfer from the polymeric insulators, as reported in the supporting information of reference \cite{Michail_2016}.

\begin{figure}
\centering
\includegraphics[width=0.9\linewidth]{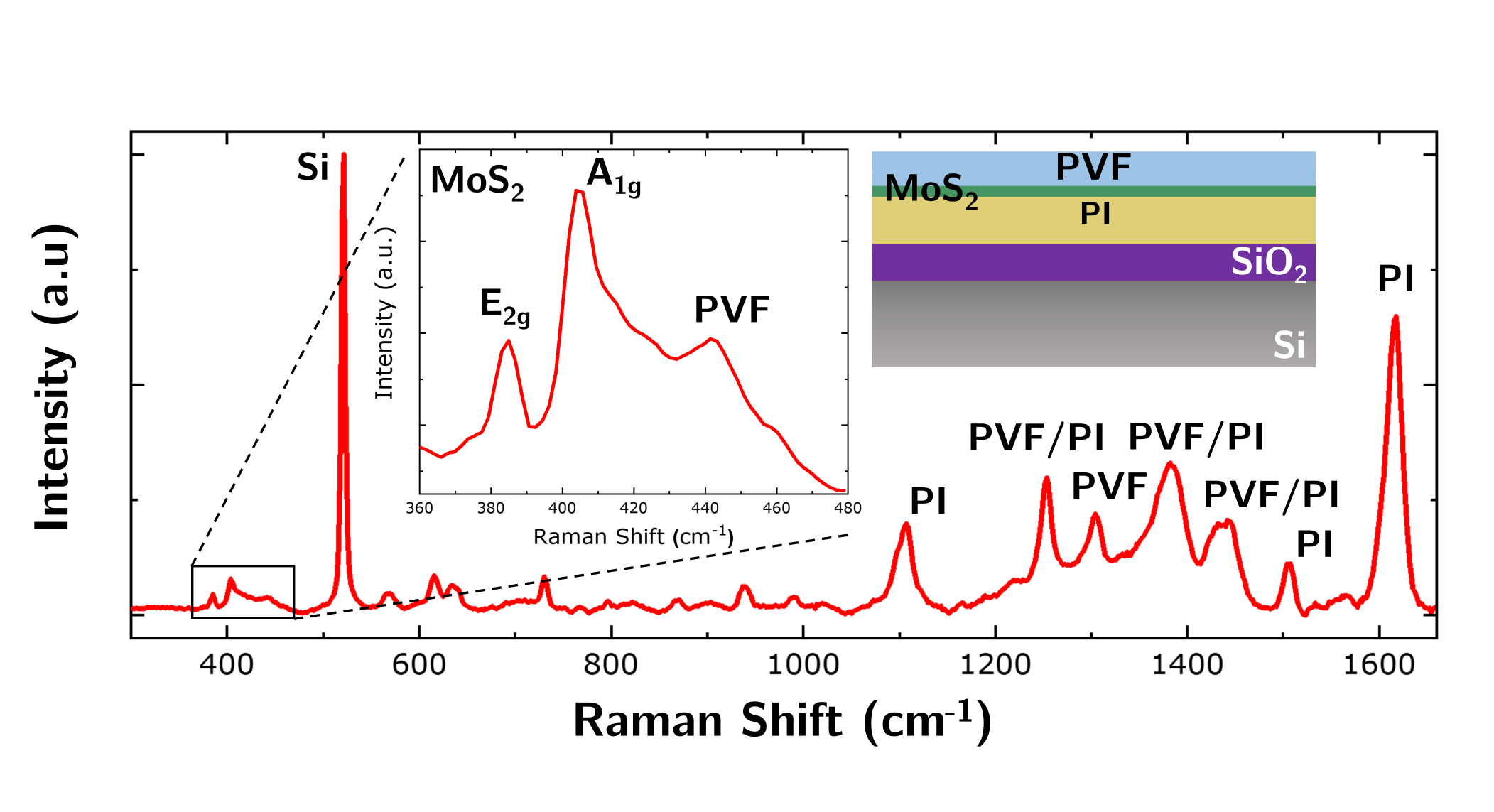}
\caption{Raman analysis of the structure PI/\ce{MoS_{2}}/PVF, attached on a \ce{SiO_{2}}/\ce{Si} substrate. The insets present: a sketch of the structure in analysis and an enlargement of the range of the \ce{MoS_{2}} Raman modes.}
\label{fig:raman_01}
\end{figure}

Figure \ref{fig:raman_01} presents the Raman spectrum of a PI/\ce{MoS_{2}}/PVF structure, fabricated on a \ce{SiO_{2}}/\ce{Si} substrate in order to have a more intense Raman signal. The sharp mode at $520~$cm$^{-1}$ is assigned to the silicon transverse optical vibrational mode. In addition, the Raman spectrum presents several vibrational modes above $1000~$cm$^{-1}$, attributed to PVF and PI. The accurate attribution of the different vibrational modes is reported in Table \ref{tab:comparison2}. The inset presents an enlargement of the range of the \ce{MoS_{2}} Raman modes, where the $E_{2g}$ and $A_{1g}$ peaks appear at $385~$cm$^{-1}$  and $405.3~$cm$^{-1}$, respectively. The Raman modes of the \ce{MoS_{2}} are superimposed to a broad band at $450~$cm$^{-1}$ assigned to the PVF. 
\begin{table}[ht]
    \fontsize{10pt}{10pt}\selectfont%
    \centering%
    \resizebox{0.9\linewidth}{!}{%
    \begin{tabularx}{0.9\linewidth}{XXXr}
        Peak Raman Shift (cm$^{-1}$) & Material & Attribution & Reference \\%
        \hline%
        1106 & PI & C--N--C transverse vibration & \cite{yang2013surface, Nishikida_2003} \\%
        1253 & PVF/PI & C--O stretching & \cite{yang2013surface, Hossain_2014, Nishikida_2003} \\%
        1305 & PVF & C-H bending vibration & \cite{yang2013surface, Hossain_2014, Nishikida_2003} \\%
        1383 & PVF/PI & C--C stretching  & \cite{yang2013surface, Hossain_2014, Nishikida_2003} \\%
        1439 & PVF/PI & C--N stretching & \cite{yang2013surface, Hossain_2014, Nishikida_2003} \\%
        1505 & PI & C=C stretching & \cite{yang2013surface, Nishikida_2003} \\%
        1616 & PI & C=C stretching  & \cite{yang2013surface, Nishikida_2003} \\%
        \hline%
    \end{tabularx}%
    }%
    \caption{The table STT presents a resume of the attribution of the Raman modes.}%
    \label{tab:comparison2}%
\end{table}

\newpage
\section{AFM measurements on FET stacks}\label{sec:AFM_char}
AFM topography and phase maps were collected using a Bruker AFM operated in the Scan assist mode. The AFM measurements were performed by scratching the surface of a dedicated device fabricated on a \ce{SiO_{2}}/\ce{Si} substrate, employing a custom in-house developed scribing system \cite{Sargeni_2024}. 

\begin{figure}
\centering
\includegraphics[width=0.9\linewidth]{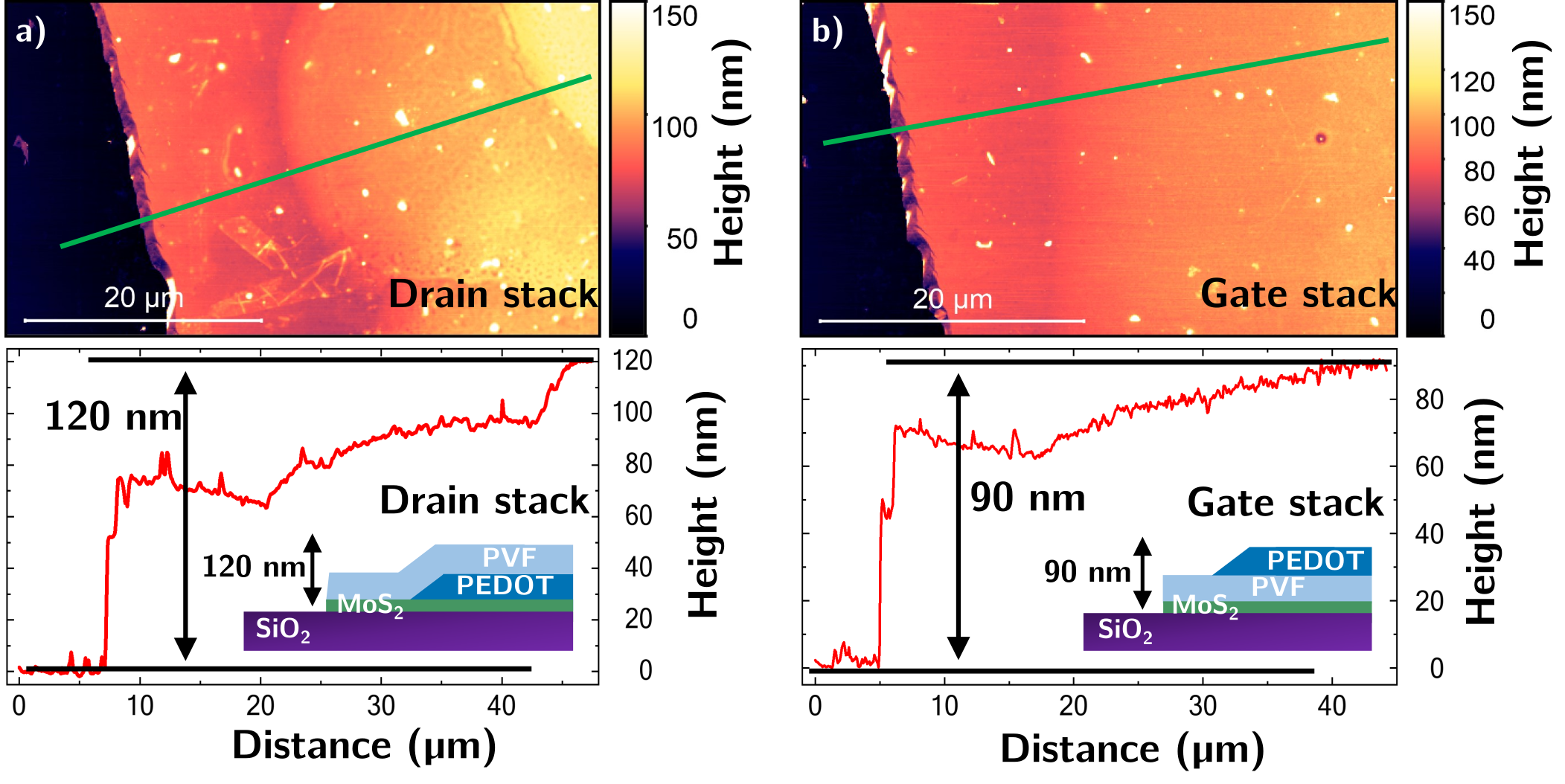}
\caption{AFM morphological maps of the drain stack (a) and the gate stack (b). The related line profiles, indicated on the map by a green dashed line, are reported below each corresponding map, with a schematic representation of the measured device stack as inset.}
\label{fig:afm_morphomaps}
\end{figure}

The AFM morphological analysis allowed an accurate evaluation of the thicknesses of both the drain stack and gate stack. Figure \ref{fig:afm_morphomaps}a reports the morphological map of the drain stack consisting of \ce{MoS_{2}}, PEDOT:PSS drain electrode and PVF, and the line profile obtained by AFM measurements, revealing a thickness of $120~$nm. A similar analysis was conducted for the gate stack (Figure \ref{fig:afm_morphomaps}b), composed of \ce{MoS_{2}}, PVF, and PEDOT:PSS gate-electrode, where the line profile reveals a total thickness of $90~$nm.

\newpage
\section{Electrical characterization and descriptive statistics}\label{sec:electrical_characterization}

The statistics of electrical parameters were derived employing a customized Python script. An ensemble of 85 transistors, sharing identical nominal dimensions of $L=65~\mu$m and $W=420~\mu$m, was characterized. For each transistor, both a forward and a backward $I_{D}$ - $V_{GS}$ curve, with $V_{GS}$ ranging from $-4~$V to $5~$V, and $V_{DS}$ set to $0.5~$V, were measured. Gate leakage ($I_{G}$) and source ($I_{S}$) currents were simultaneously measured too.

The script processed raw data from individual devices within the dataset, excluding any defective units (e.g., those exhibiting short circuits between electrode pairs or open circuits between source and drain) from analysis. For every acceptable device, the $I_{D}$ - $V_{GS}$ curves were further processed. To reduce the impact of noise, each curve was smoothed with a least-squares polynomial fitting. Two separate sets of parameters (from the forward and backward curve) were extracted for each device.

\begin{figure}
\centering
\includegraphics[width=0.9\linewidth]{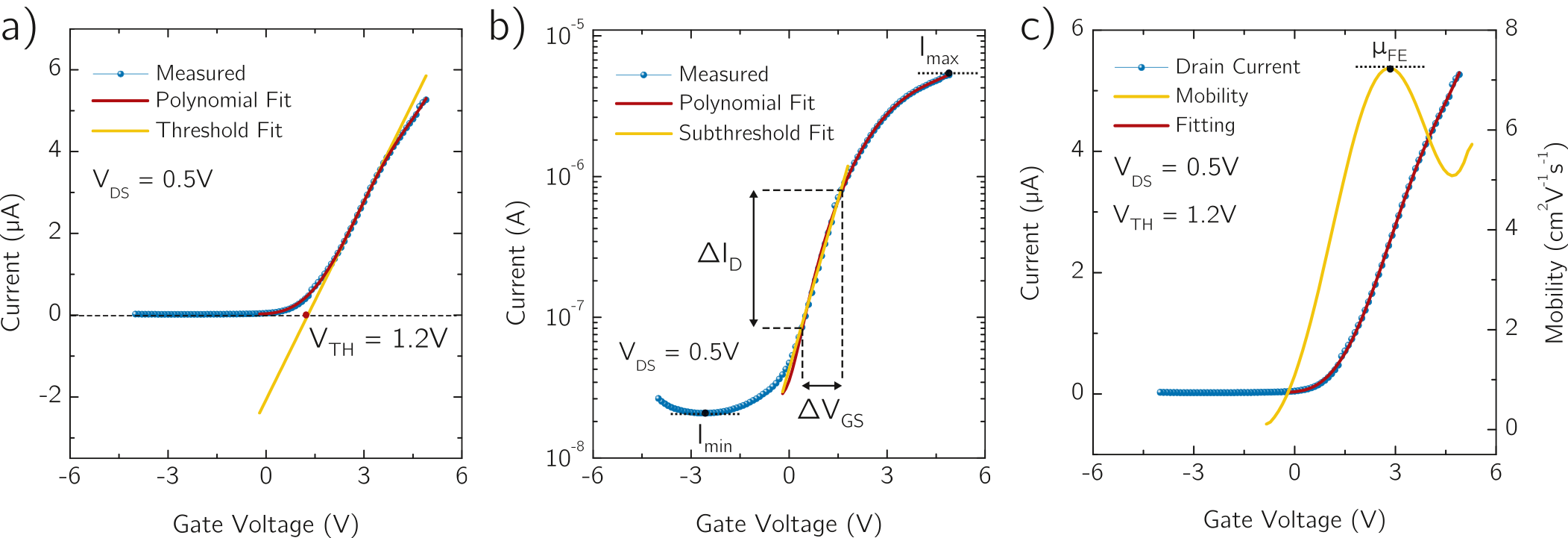}
\caption{Graphical representation of the parameters extraction process. a) Polynomial fitting (red line) of a target transistor's transfer characteristic curve (blue line). The threshold voltage value is determined by the intercept point where the tangent to the transfer curve (yellow line), at its maximum slope, intersects the $V_{GS}$ axis. b) Polynomial fitting (red line) of the same transfer characteristic curve (blue line), depicted in semi-logarithmic scale. The subthreshold fit line (yellow) is utilized to determine the boundaries of the $V_{GS}$ interval within which the transfer curve experiences a tenfold increase in $I_{DS}$, specifically within the subthreshold region ($V_{GS}<V_{TH}$). Additionally, $I_{max}$ and $I_{min}$ are indicated on the transfer curve. c) Representation of field effect mobility (yellow line) as a function of the gate voltage, calculated from the same transfer characteristic (blue line). The target mobility value is extracted in its maximum.}
\label{fig:meas_analysis}
\end{figure}

Threshold voltages ($V_{TH}$) were determined as the $V_{GS}$ value where the tangent line to the $I_{D}$ - $V_{GS}$ curve, at its maximum slope intersects the $V_{GS}$ axis (i.e., with the maximum transconductance method). A graphical representation of the procedure can be found in Figure \ref{fig:meas_analysis}a.

The values of $V_{TH}$ were subsequently used for the evaluation of the field-effect mobility ($\mu_{FE}$) values. To this purpose, we used the expressions derived for both the linear and the saturation regime of an ideal long-channel MOSFET for $V_{GS}>V_{TH}$:
\begin{equation*}\label{eq:mobility2}
\mu_{FE} = \genfrac{\{}{.}{0pt}{}{\frac{L}{W}\frac{1}{C_i}\frac{1}{V_{DS}}\frac{\partial{I_D}}{\partial{V_{GS}}}, \textrm V_{DS}<V_{GS}-V_{TH}}{2\frac{L}{W}\frac{1}{C_i}(\frac{\partial{\sqrt{I_D}}}{\partial{V_{GS}}})^2, \textrm V_{DS}>V_{GS}-V_{TH}}
\end{equation*}
where $C_{i}$ is the insulator film capacitance per unit area.
An example of mobility value extraction can be found in Figure \ref{fig:meas_analysis}c.

Regarding the on-off current ratios, \cite{Cheng_2022} proposed an alternative metric to the traditional $I_{on}$/$I_{off}$ ratios, due to the ambiguity in the choice of the gate voltages corresponding to the $ON$ and $OFF$ states. We thus adopted their methodology and computed the $I_{max}$/$I_{min}$ ratios, evaluated within the linear regime, as a more robust indicator of FET performance for digital applications.

Finally, the subthreshold slope ($SS$) values were computed as the minimum change in $\Delta V_{GS}$ required to achieve a tenfold increase in $I_{DS}$, specifically within the exponential regime of the transfer curves (i.e., in the subthreshold regions, where $V_{GS}<V_{TH}$). Figure \ref{fig:meas_analysis}b illustrates the procedure.

\begin{figure}
\centering
\includegraphics[width=0.9\linewidth]{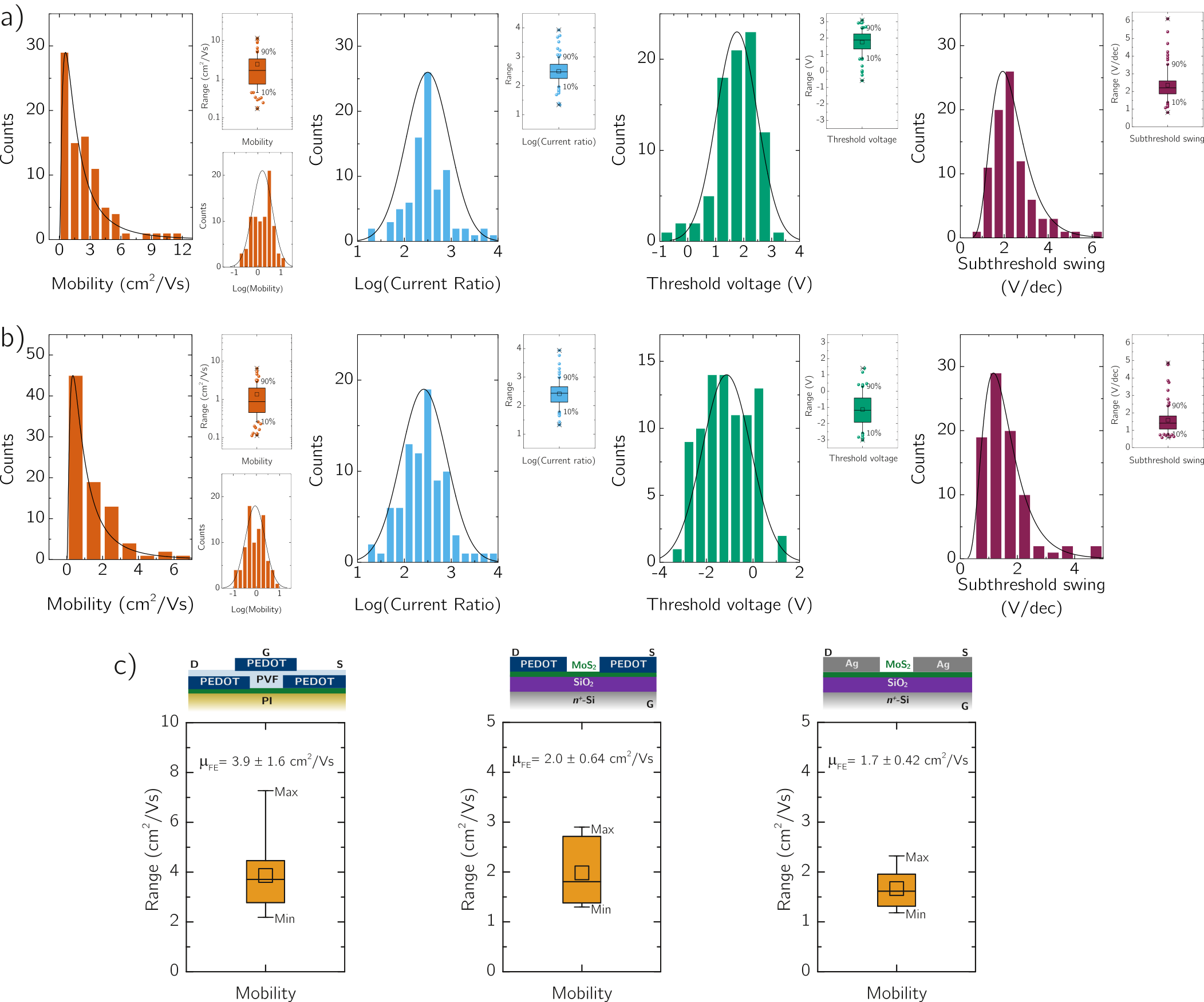}
\caption{Hystograms reporting the statistical distributions of the forward biasing (a) and backward biasing (b) electrical parameters, calculated for the characterized set of 85 working FETs fabricated on the same flexible PI substrate, with PVF as gate dielectric and PEDOT:PSS as conductor. Mobility and subthreshold swing are fitted against a log-normal distribution, while threshold voltage and logarithm of current ratio against a normal distribution. The box representations, featured in the insets, show the mean and median values, while spanning from the 25$^\textrm{th}$ to the 75$^\textrm{th}$ percentiles. Additionally, mobility is shown logarithmically transformed in the insets and fitted against a normal distribution. \textcolor{black}{c) Comparison among the values of mobility extracted from the characterization of the \ce{MoS_{2}} films on different FET structures. On the left, our printed PEDOT:PSS and PVF-based FETs, fabricated on a flexible PI substrate; in the middle and on the right, two back-gate configurations where a highly doped \ce{Si}/\ce{SiO_{2}} wafer has been used as the substrate for the transferred \ce{MoS_{2}}, but two different materials have been used for the top contacts (PEDOT:PSS, in the middle, and Ag, on the right). The box representations show the mean and median values from the 25$^\textrm{th}$ to the 75$^\textrm{th}$ percentiles. The mobility is reported within each graph, represented with its mean and standard deviation values.}}
\label{fig:stats}
\end{figure}

Statistical distributions and complete descriptive statistics of the forward and backward parameters, extracted with the above procedures, can be found in Figure \ref{fig:stats} and in Table \ref{tab:descriptive_statistics}. Values of mobility ($\mu_{FE}$), threshold voltage ($V_{TH}$), subthreshold swing ($SS$) and current ratio ($I_{max}$/$I_{min}$) are reported.

\begin{table}
    \fontsize{8pt}{8pt}\selectfont%
    \centering%
    \resizebox{0.9\linewidth}{!}{%
    \begin{tabularx}{0.9\linewidth}{XXXXXXXXr}%
        & Bias & Mean & Std. dev. & L $95\%$ CI & U $95\%$ CI & Min & Med & Max \\%
        \hline%
        $\mu_{FE}$%
        & fw & 2.44 & 2.30 & 1.94 & 2.93 & 0.172 & 1.68 & 11.5 \\%
        (cm$^{2}$/Vs)%
        & bw & 1.35 & 1.27 & 1.08 & 1.63 & 0.112 & 0.865 & 6.45 \\ \\%
        $V_{TH}$ %
        & fw & 1.76 & 0.734 & 1.60 & 1.92 & -0.575 & 1.89 & 3.10 \\%
        (V)%
        & bw & -1.13 & 1.04 & -1.36 & -0.909 & -3.02 & -1.17 & 1.41 \\ \\%
        $SS$%
        & fw & 2.36	& 0.930 & 2.16 & 2.56 & 0.817 & 2.21 & 6.12 \\%
        (V/dec)%
        & bw & 1.58 & 0.823 & 1.41 & 1.76 & 0.584 & 1.41 & 4.86 \\ \\%
        $I_{max}$ %
        & fw & 1.92 & 1.85 & 1.52 & 2.32 & 0.111 & 1.24 & 9.64 \\%
        ($\mu$A)%
        & bw & 2.01 & 1.93 & 1.59 & 2.427 & 0.121 & 1.31 & 10.1 \\ \\%
        $I_{min}$ %
        & fw & 9.18 & 13.8 & 6.20 & 12.2 & 0.239 & 5.06 & 95.3 \\%
        (nA)%
        & bw & 12.8 & 21.2 & 8.18 & 17.3 & 0.311 & 6.20 & 123 \\ \\%
        $I_{max}$/$I_{min}$%
        & fw & 631 & 1217 & 369 & 894 & 21.7 & 302 & 8523 \\%
        & bw & 554 & 1165 & 303 & 806 & 21.2 & 272 & 8744 \\%
        \hline%
    \end{tabularx}%
    }%
    \caption{Descriptive statistics for the characterized array of 85 working transistors fabricated on the same flexible PI substrate. Mean, standard deviation, upper and lower $95\%$ confidence intervals, minimum, maximum and median values are reported for extracted mobility ($\mu_{FE}$), threshold voltage ($V_{TH}$), subthreshold swing ($SS$), maximum ($I_{max}$) and minimum ($I_{min}$) drain currents and their ratio ($I_{max}$/$I_{min}$). Values are extracted from the set of measured transfer characteristics. Entries labeled as "fw" and "bw" refer to the forward (rising $V_{GS}$) and backward (falling $V_{GS}$) biasing, respectively.}%
    \label{tab:descriptive_statistics}%
\end{table}
\FloatBarrier
\textcolor{black}{Despite the good field-effect mobility values resulting from the characterization of our devices, the obtained average mobility is lower than others in the literature involving CVD and MOCVD processes. 
To evaluate the influence of each of the above-mentioned device materials on the mobility, we have performed a comparative analysis on the performance of FETs, considering three different device structures as the ones shown in Figure \ref{fig:stats}c, i.e., a top-gate and a bottom-gate configuration  fabricated on a highly doped \ce{Si}/\ce{SiO_{2}} wafer, based both on transferred MOCVD \ce{MoS_{2}} as semiconductor, and  two different inks, PEDOT:PSS and Ag, as conductive materials for the inkjet-printed top electrodes (source/drain) in the back-gate configuration. All the devices share identical nominal dimensions of $L=100~\mu$m and $W=500~\mu$m. The thickness of the double-layer PVF structure is about $50~$nm and the \ce{SiO_{2}} is $285~$nm thick. The electron mobility has been extracted, following the same procedures we adopted  and described in the manuscript, resulting to be about $3.9$, $2$ and 1.7~cm$^{2}$V$^{-1}$s$^{-1}$ for, respectively, the top-gate configuration and the two back-gate configurations. From this analysis, we see that the measurements on back-gate configurations lead to average mobility values of the same order of magnitude as the ones we obtained with our top-gate devices fabricated on PI with PVF as a gate.}

\newpage
\section{Ultra-thin capacitors fabrication and characterization}\label{sec:capacitors_fabrication_char}
\begin{figure}
\centering
\includegraphics[width=0.86\linewidth]{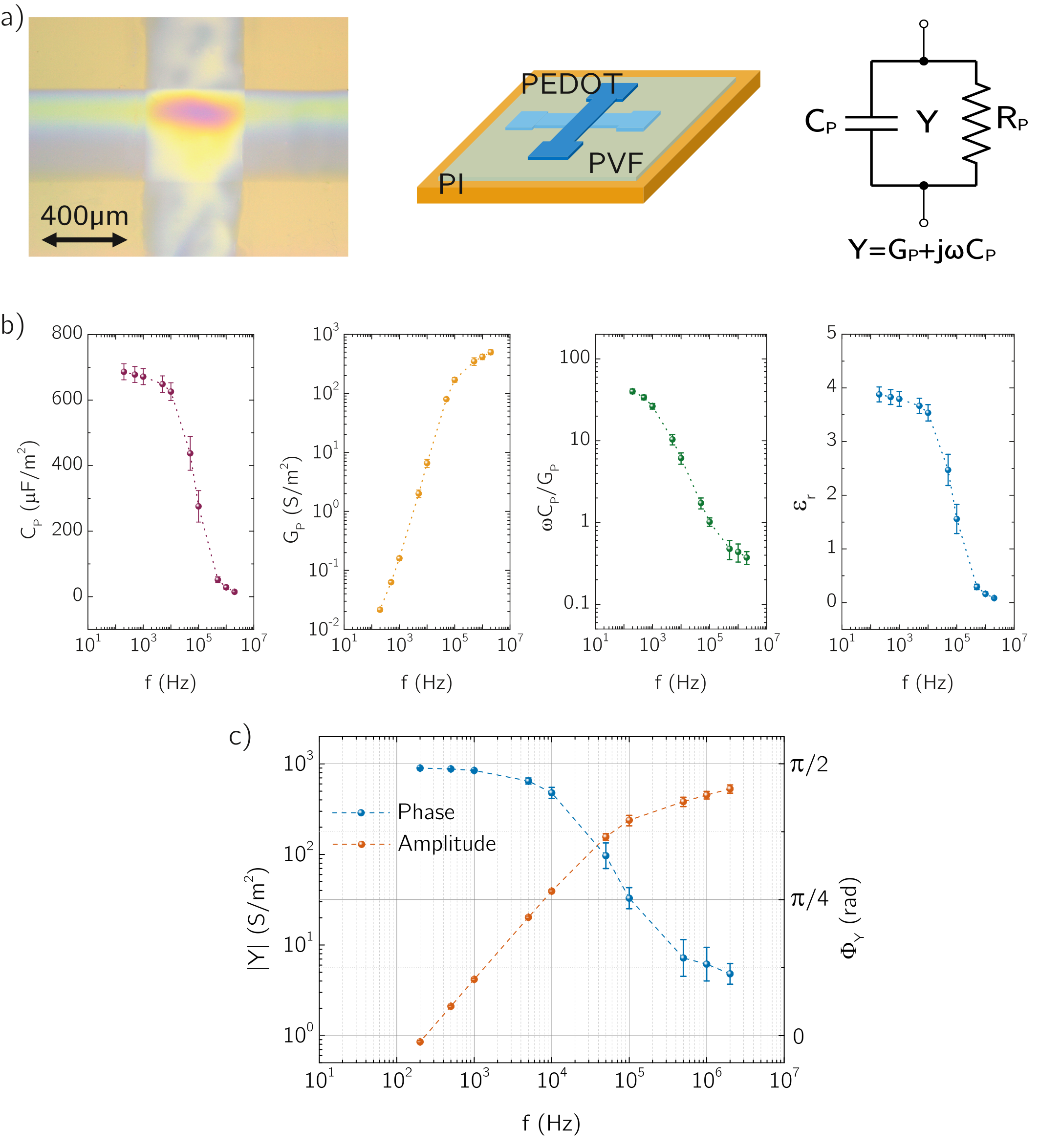}
\caption{a) Optical micrograph, sketch and equivalent electric schematic of a parallel plates capacitor fabricated on PI, with top/bottom electrodes made of PEDOT:PSS, and $50~nm$ thick PVF as dielectric material. b) Dynamic electrical characterization of an array of $24$ working capacitors fabricated on the same PI substrate. The capacitance per unit area $C_{P}$, the conductance per unit area $G_{P}$, the admittance ratio $\omega C_{P}/G_{P}$, and the relative permittivity $\epsilon_{r}$ are reported as functions of the applied voltage frequency, spanning from $200~$Hz to $2~$MHz. Mean values are reported and error bars represent the standard deviation. C) Amplitude ($|Y|$) and phase ($\Phi_{Y}$) representation of the mean admittance $Y=G_{P}+j\omega C_{P}$ as a function of frequency, for the same set of capacitors. The standard deviation is represented by error bars.}
\label{fig:capacitors_characterization}
\end{figure}

Parallel plate capacitors featuring inkjet-printed PEDOT:PSS as top/bottom electrodes and PVF as insulator were fabricated on PI substrates to assess the electrical properties of PVF thin films, as illustrated in Figure \ref{fig:capacitors_characterization}a. The fabrication process replicates that of FETs, with the omission of the step involving \ce{MoS_{2}} transferring. 

A set of $24$ working capacitors fabricated on the same PI substrate was characterized by modeling the capacitors with an $R_{P}$--$C_{P}$ parallel model, wherein $R_{P}$ accounts for leakage through the dielectric film phenomena. The impact of series resistance was disregarded, given the high conductivity of the PEDOT:PSS electrodes.
$C_{P}$ and $R_{P}$ were estimated by evaluating the capacitor performance across a frequency range, from the DC to the high-frequency regime. The measurements were conducted using a four-probe configuration while applying a $1~$V amplitude sinusoidal signal. An open circuit calibration was performed before data acquisition to evaluate the system's parasitic capacitance, considered as an offset in the capacitance measurements. 

Figure \ref{fig:capacitors_characterization}b reports the average values of the capacitance per unit area $C_{P}$, the conductance per unit area $G_{P}$, the admittance ratio $\omega C_{P}/G_{P}$ between the capacitive susceptance and the parallel parasitic conductance, and the relative permittivity $\epsilon_{r}$, for the characterized set of capacitors, with error bars representing the standard deviation. 
The $\omega C_{P}/G_{P}$ ratio is a crucial figure of merit for real capacitors, indicating their quality in terms both of material properties and fabrication reliability. A higher value signifies a device closer to its ideality, where the leakage phenomena through the insulating layer can be neglected. Our capacitors exhibit exceptional reliability with a constant value of $\epsilon_{r}$ up to frequencies on the order of tens of kilohertz, enabling their application in a wide range of digital and analog electronics.

The dielectric constant was derived from the measured capacitance values according to the parallel plate capacitor model: 
\begin{equation*}\label{eq:relative_permittivity}
\epsilon_{r} = \frac{C_{P} t}{A \epsilon_{0}}
\end{equation*}
where $A$ is the area of the capacitor (designed to be a square with a side of $400~\mu$m), $\epsilon_{0}$ is the vacuum permittivity value ($8.85\cdot 10^{-12}~$AsV$^{-1}$m$^{-1}$, while an average thickness ($t$) equal to $50~$nm of the PVF insulating layer was considered.

A representation of the mean admittance $Y=G_{P}+j\omega C_{P}$, in terms of both amplitude ($|Y|$) and phase ($\Phi_{Y}$), as a function of frequency is depicted in Figure \ref{fig:capacitors_characterization}c for the same set of capacitors, with error bars representing the standard deviation of values. Consistent with the conclusions regarding the $\omega C_{P}/G_{P}$ ratio, the graph illustrates that the phase of the admittance maintains a value close to $\pi /2$ for frequencies up to $10^{4}~$Hz, indicating its nearly ideal capacitive behavior with negligible parasitic effects.

\newpage
\section{Electromechanical characterization}\label{sec:electromech_char}
The pliability and conformability of PEDOT:PSS-PVF parallel plate capacitors and \ce{MoS_{2}}-PEDOT:PSS-PVF transistors were evaluated under both static and dynamic bending conditions. 

Static characterization involved examining their electrical response under varying static bending conditions with diverse curvature radii \textcolor{black}{from $14~$mm to $2~$mm}. 
\textcolor{black}{The corresponding strain values were calculated following the bending-induced strain formula:
\begin{equation*}\label{eq:strain}
\epsilon = \frac{t}{2R}
\end{equation*}
where $\epsilon$ is the strain, $t$ is the device thickness and $R$ is the radius of curvature.}

\begin{figure}
\centering
\includegraphics[width=0.9\linewidth]{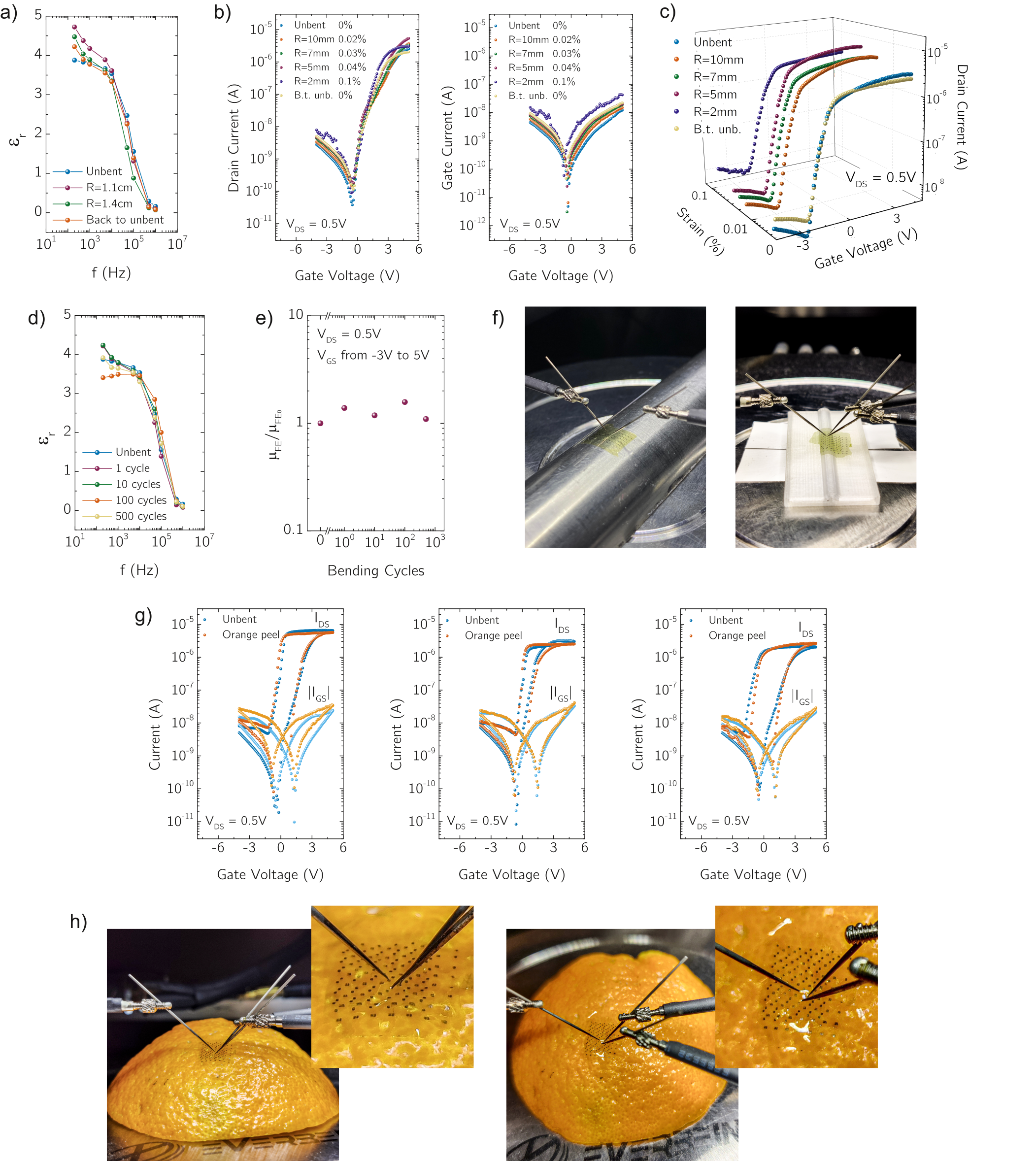}
\caption{\textcolor{black}{Characterization of our devices under static bending conditions, for different bending radii: a) Relative permittivity curves of PVF ($\epsilon_{r}$) as functions of the frequency. b) Transfer curves (drain currents on the left and gate currents on the right) of FETs. c) Transfer curves of FETs as a function of the applied strain. Characterization of our devices under repeated bending cycles: d) Relative permittivity curves of PVF as function of the frequency; e) field-effect mobility $\mu_{FE}$ of FETs, normalized to their starting value $\mu_{FE_{0}}$ (before undergoing mechanical stress), as a function of the bending cycles. f) Micrograph of the measurement setup for static characterization. Characterization of our devices transferred on an irregular surface:
g) Transfer curves for FETs conformed to an orange peel (h), compared with the starting unbent condition.}}
\label{fig:bending_characterization}
\end{figure}

Dynamic characterization aimed to demonstrate their functionality during repeated bending cycles, with up to $500$ cycles performed. 
The results of this investigation are depicted in Figure \ref{fig:bending_characterization}.

Capacitors were characterized by determining the average relative permittivity ($\epsilon_{r}$) as a function of the applied voltage frequency, ranging from $200~$Hz to $2~$MHz. Transistors were characterized by measuring their transfer characteristic ($I_{D}$ as a function of $V_{GS}$) at a fixed $V_{DS}$ value of $0.5~$V, while concurrently extracting their average mobility ($\mu_{FE}$) as a control parameter. 
Negligible alterations in their electrical behavior were observed during both static and dynamic characterizations. Any minor variations can be attributed to the measurement conditions themselves, as applying the probes becomes more challenging when the device is bent, potentially compromising the measurement accuracy. 

\textcolor{black}{We have also performed an electrical characterization of the \ce{MoS_{2}}-PEDOT:PSS-PVF transistors transferred on an orange peel, in order to prove their functionality while adhering to irregular, rough and bent surfaces. Figure \ref{fig:bending_characterization}h shows the measurement setup and Figure \ref{fig:bending_characterization}g the trans-characteristic curves of three different devices, directly characterized while adhering on the orange peel. These characteristics are compared with the ones obtained in a flat state. The curves show that our devices operate correctly when adhering to complex surfaces and demonstrate robustness to mechanical stress, which meets the requirements of conformable electronics applications.}

\begin{figure}
\centering
\includegraphics[width=0.6\linewidth]{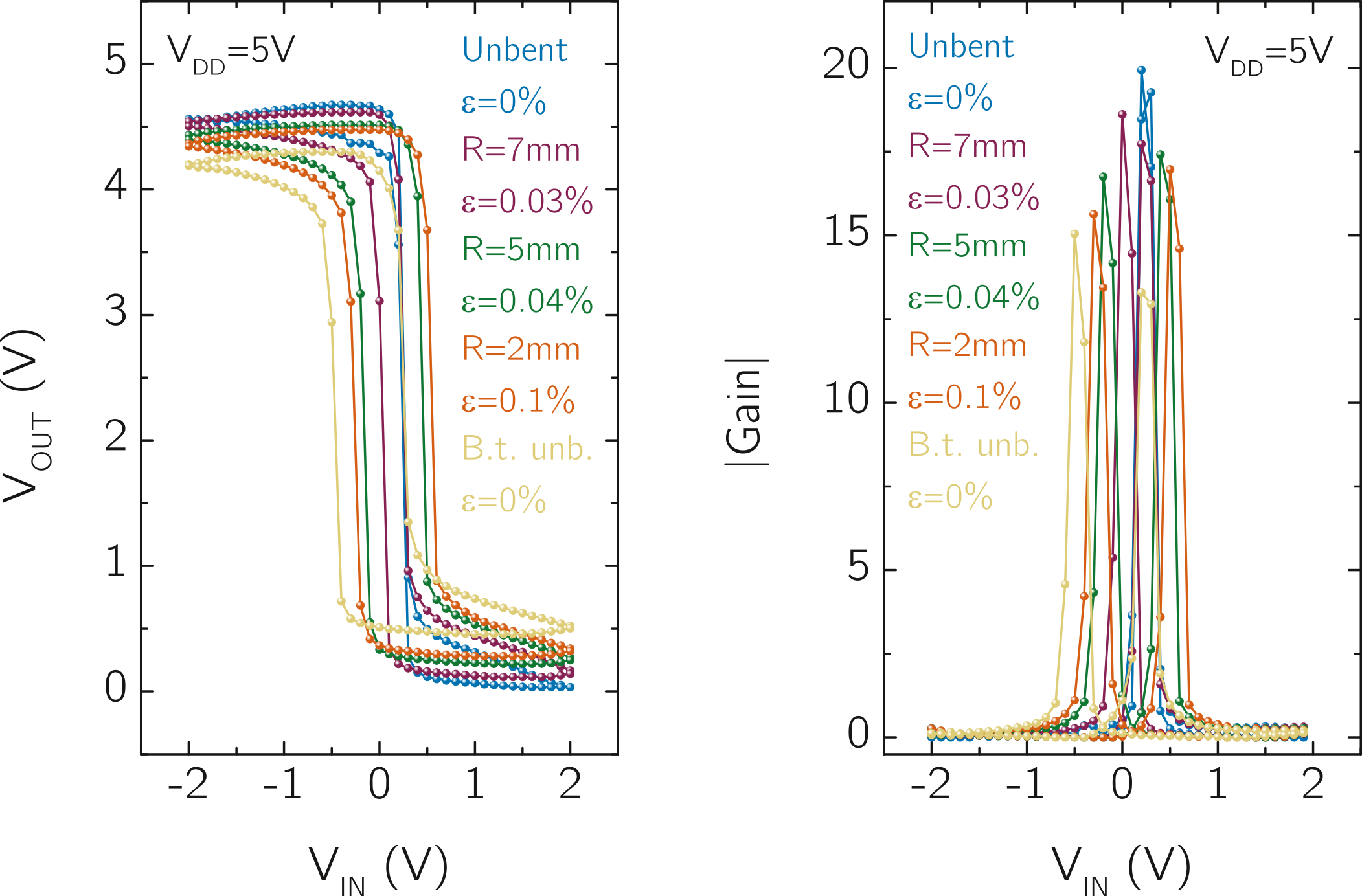}
\caption{\textcolor{black}{Characterization of our circuits under static bending conditions, for different bending radii: transfer characteristic (left) and voltage gain modulus (right), as functions of the input gate voltage, of a depletion-load inverter gate, measured for a supply voltage of 5 V}}
\label{fig:bending_inverter}
\end{figure}

\textcolor{black}{Finally, to assess the functionality of our circuits under bending conditions, we characterized an inverter gate evaluating its transfer characteristic and its voltage gain while undergoing bending-induced stresses.}
\textcolor{black}{The results of this study are shown in Figure \ref{fig:bending_inverter}, confirming that the circuit maintains its functionality as an inverter while undergoing mechanical stresses, as required by conformable electronics applications.}

\newpage
\section{Table of comparison}\label{par:comparison}
A comparative analysis of various studies on flexible and conformable transistors and circuits reported in the recent literature, focusing on parameters such as transistor film thickness, utilized materials, fabrication techniques and electrical performance (e.g., mobility, operating voltage, and threshold voltage), is reported in Table \ref{tab:comparison}. The operating voltage refers to the gate voltage value for which the field-effect mobility has been calculated, when reported.
\begin{landscape} 
\thispagestyle{empty}
\begin{table}[htb]
    \fontsize{4pt}{5pt}\selectfont
    \centering
    \begin{tabularx}{1.0\linewidth}{ X X X X X X X X  X  X X X  X r }
    \hline
        & Thickness & Gate stack & Substrate & S, D electrodes & G electrode  & Semiconductor & Dielectric & Technology & Op. voltage & Mobility & Th. voltage & Strengths & Limitations \\%
        & ($\mu$m) & (nm) & ($\mu$m) & (nm) & (nm) & (nm) & (nm) &  & (V) & (cm$^{2}$/Vs) & (V) &  & \\%
         \hline
        This work & $4$ & $90$ & PI, $3.8$ & PEDOT, 60 & PEDOT, $60$ & s.l. \ce{MoS_{2}}, $<1$ & PVF, $50$ & Sol., MOCVD & $5$ & $2.44$ & $1.76$ & Op. voltage, Low-budget, Circuits & \ce{MoS_{2}} transfer \\%
        \cite{Zhao_2016} & $300$ & $60$ & PET, $300$ & \ce{Ti}/\ce{Au}, $2$/$30$ & \ce{Au}, $30$ & s.l. \ce{MoS_{2}}, $<1$ & \ce{HfO_{2}}, $30$ & Sol., CVD, ALD, Litho & $8$ & $13.9$ & N.A. & Mobility, Op. volt. & Masks, HF, Common gate \\%
        \cite{Mirshojaeian_2023} & $3$ & $120$ & Parylene, $0.06$ & \ce{Au}, $30$ & \ce{Au}, $30$ & DNTT, $30$ \textit{or} PDI-8CN2, $30$ & Parylene, $60$ & Sol., CVD, Th. dep. & 5 & 0.11 & 1 & Volt., CMOS & Masks, Comm. gate, Mob. \\%
        \cite{Viola_2021} & $0.15$ & $130$ & Freestanding & PEDOT, $20$ & PEDOT, $60$ & DPP-TVT, $20$ & PVF, $50$ & Sol. & $6$ & $0.092$ & $0.9$ & Fully sol. based, Op. volt. & Mobility \\%
        \cite{Liu_2018} & $0.59$ & $560$ & Freestanding & \ce{Au}, $30$ & \ce{Au}, $30$ & PTCDI-C1, $30$ & PMMA/PVA, $500$  & Sol., Th. dep. & $60$ & $0.52$ & N.A. &  & Comm. gate, Masks \\%
        \cite{Ren_2018} & $0.38$ & $360$ & Freestanding & \ce{Au}, $20$ & \ce{Au}, $20$ & C8-BTBT, $20$ & C-PVA, $320$ & Th. dep. & $60$ & $7.22$ & $-15$ & Mobility & Comm. gate, Masks \\%
        \cite{Fukuda_2016} & $0.4$ & $350$ & Freestanding & \ce{Au}, $50$ & \ce{Au}, $50$ & DNTT, $50$ & Parylene-SR, $250$ & Sol., Th. dep., PVD  & $10$ & $0.37$ & $-0.55$ & Op. volt. & Masks, Mobility \\%
        \cite{Jang_2019} & $0.5$ & $465$ & Freestanding & \ce{Au}, $35$ & \ce{Au}, $20$ & Pentacene, $30$ & PVDF-TrFE, $415$ & Sol., EBD, Th. dep. & $30$ & N.A. & N.A. &  & Op. volt., Masks \\%
        \cite{Nawrocki_2016} & $0.3$ & $124$ & Parylene, $0.060$ & \ce{Au}, $30$ & \ce{Au}, $30$ & DNTT, $30$ & Parylene, $64$ & Sol., CVD, Th. dep. & $5$ & $0.34$ & $-1.72$ & Op. volt. & \\%
        \cite{Stucchi_2020} & $>2.3$ & $>220$ & Parylene, $2$ & PEDOT, $40$ & PEDOT, $40$ & DPP-TT \textit{and} P(NDI2OD-T2), N.A. & PMMA, $20$ \& Parylene, $160$ & Sol., CVD & $10$ & $0.1$ & N.A. & CMOS, Op. volt., & Mobility, \\%
        \cite{Lai_2020} & $>0.950$ & $>250$ & Parylene-C, $0.7$ & PEDOT, N.A. & PEDOT, N.A. & TIPS Pentacene, N.A. & Parylene-C, $250$ & Sol., Th. Dep., CVD & $7$ & $0.13$ & $0$ & Op. volt. & Mobility \\%
        \cite{Lai_2017} & $>0.630$ & $>190$ & Parylene, $0.4$ & \ce{Au}, $40$ & \ce{Al}, $40$ & TIPS Pentacene \& ActiveInk, N.A. & \ce{Al_{2}O{3}}/Parylene-C, $20$/$130$ & Sol., Th. dep., Litho., CVD & $6$ & $0.14$ & $0.65$ & Op. volt, CMOS, High freq. & Mobility \\%
        \cite{Zhao_2021} & $10$ & $795$ & PDMS, $>9$ & \ce{Au}, $30$ & \ce{Au}, $30$ & DNTT, $50$ & PVA/DC1–2577, $285$/$430$ & Litho., Sol., Plasma & $80$ & $0.68$ & $-0.51$ & High density & Op. volt., Masks \\%
        \cite{Mun_2023} & $5$ & $120$ & Parylene, $4$ & \ce{MoO_{3}}/\ce{Au}, $30$/$40$ & \ce{Al}, $45$ & \ce{Ph-BTBT-10}, $30$ & pV3D3, $45$ & Th. Dep., CVD, Sol. & $2.5$ & $4.5$ & $0$ & Op. volt. & Masks \\%
        \cite{LiuM_2018} & $0.830$ & $450$ & PVA, $0.35$ & \ce{Au}, $30$ & \ce{Au}, $30$ & \ce{PTCDI-C_{13}}, $40$ & PMMA, $380$ & Sol., Th. Dep., Litho & $70$ & $0.58$ & N.A. &  & Op. volt., Masks \\%
        \cite{munzenrieder_2015} & $1.2$& $110$ & Parylene, $1$ & \ce{Ti}/\ce{Au}, $10$/$75$ & \ce{Cr}, $35$ & IGZO, $50$ \textit{or} \ce{NiO}, $50$ & \ce{Al_{2}O_{3}}, $25$ & Th. Dep., EBD, Sput., Litho & $5$ & $11$ & $0.4$ & Mobility, Op. volt., CMOS & Masks \\%
        \cite{Hoang_2023} & $1.6$ & $63$ & PI, $1.5$ & \ce{Cr}/\ce{Au}, $3$/$30$ & \ce{Cr}/\ce{Au}, $3$/$30$ & sl. \ce{MoS_{2}}, $<1$ & \ce{Al_{2}O_{3}}, $30$ & CVD, ALD, Th. dep & $10$ & $6.5$ & $3.8$ & Mobility, Op. volt. & Complex \\%
        \cite{Jongwon_2013} & N.A. & $>400$ & PET, N.A. & Graphene, N.A. & Graphene, N.A. & exf. m.l. \ce{MoS_{2}}, N.A. & c-PVP, $400$ & Sol., CVD, RIE, Litho. & $20$ & $0.7$ & $27.1$ &  & Masks, Op. volt. \\%
        \cite{Reato_2022} & $8.2$ & $170$ & PI, $8$ & \ce{Ni}, $50$ & \ce{Al}/\ce{Ti}, $100$/$35$ & s.l. \ce{MoS_{2}}/\ce{h-BN},  $<2$ & \ce{Al_{2}O_{3}}, $35$ & MOCVD, Sol., Litho., ALD, RIE, Sput. & $10$ & $1.8$ & $-4.4$ &  & Masks, Complex \\%
        \cite{Wang_2018} & $100$ & $>1380$ & SEBS, N.A. & CNT, N.A. & CNT, N.A. & 29-DPP-SVS-(2), $130$ & SEBS-X-azide, $1250$ & Sol., Th. dep., Litho. & $30$ & $1.37$ & $-1$ & High density & Op. volt., Masks, Solvents \\%
        \cite{Gong_2016} & $5$ & $130$ & PI, $4.8$ & \ce{Au}, $50$ & \ce{Ti}, $100$ & s.l. \ce{WS_{2}}, $<1$ & \ce{Al_{2}O_{3}}, $30$ & Sol., CVD, Th dep., ALD, Sput., Litho.  & $12$ & $2$ & N.A. & Mobility, Op. volt. & Complex, Masks \\%
        \cite{LI_2020} & \textcolor{black}{N.A.} & \textcolor{black}{$65$} & \textcolor{black}{PET, N.A} & \textcolor{black}{\ce{Au}/\ce{Ti}, $33$/$2$} & \textcolor{black}{\ce{ITO}, $30$} & \textcolor{black}{s.l. \ce{MoS_{2}}, $<1$} & \textcolor{black}{\ce{Al_{2}O_{3}}, $35$} & \textcolor{black}{Sol., CVD, Th dep., ALD, Sput., Litho.}  & \textcolor{black}{$20$} & \textcolor{black}{$55$} & \textcolor{black}{$1$} & \textcolor{black}{Mobility, High density, Circuits} & \textcolor{black}{Complex, Masks, Op. Voltage} \\%
        \cite{ZAN_2024} & \textcolor{black}{N.A.} & \textcolor{black}{$59$} & \textcolor{black}{Dragon Skin$^{TM}$/\ce{HfO_{x}}, N.A./$0.011$} & \textcolor{black}{d.l. \ce{Graphene}, $2$} & \textcolor{black}{d.l. \ce{Graphene}, $2$} & \textcolor{black}{s.l. \ce{MoS_{2}}, $<1$} & \textcolor{black}{\ce{HfO_{x}}/cross-linked \ce{PR}, $16$/$28$} & \textcolor{black}{Sol., CVD, Th dep., ALD, Plasma, Litho.}  & \textcolor{black}{$5$} & \textcolor{black}{$4.9$} & \textcolor{black}{$2.8$} & \textcolor{black}{Mobility, Op. volt.} & \textcolor{black}{Complex, Masks} \\%
        \cite{TANG_2023} & \textcolor{black}{$120$} & \textcolor{black}{$18$} & \textcolor{black}{PET, N.A.} & \textcolor{black}{\ce{Au}, $20$} & \textcolor{black}{\ce{Au}/\ce{Ti}, $5$/$2$} & \textcolor{black}{s.l. \ce{MoS_{2}}, $<1$} & \textcolor{black}{\ce{HfO_{2}}, $10$} & \textcolor{black}{Sol., CVD, E-beam evap., RIE, ALD, Litho., Plasma}  & \textcolor{black}{$7$} & \textcolor{black}{$68$} & \textcolor{black}{$1$} & \textcolor{black}{Mobility, Op. volt., Circuits} & \textcolor{black}{Complex, Masks} \\%
        \hline%
    \end{tabularx}
    \caption{}%
    \label{tab:comparison}%
\end{table}
\end{landscape}

\newpage
\bibliography{Bibliography}

\end{document}